\documentclass[11pt]{article}
\usepackage{amssymb,amsmath,amsfonts}
\usepackage{graphicx}
\usepackage{graphics}
\usepackage{eepic,epsfig}
\usepackage{dcolumn}

\@addtoreset{equation}{section}

\makeatother

\begin{document}

\title{Electromagnetic field correlators and the Casimir effect\\
for planar boundaries in AdS spacetime\\
with application in braneworlds }
\author{A. A. Saharian,\thinspace\ A. S. Kotanjyan,\thinspace\ H. G.
Sargsyan \\
\\
\textit{Department of Physics, Yerevan State University,}\\
\textit{1 Alex Manoogian Street, 0025 Yerevan, Armenia}}
\maketitle

\begin{abstract}
We evaluate the correlators for the vector potential and for the field
strength tensor of the electromagnetic field in the geometry of two parallel
planar plates in AdS spacetime. Two types of boundary conditions are
considered on the plates. The first one is a generalization of perfect
conductor boundary condition and the second one corresponds to the confining
boundary conditions. By using the expressions for the correlators, the
vacuum expectation values (VEVs) of the photon condensate and of the electric
and magnetic fields squared are investigated. As another important local
characteristic of the vacuum state we consider the VEV of the
energy-momentum tensor. The Casimir forces acting on the plates are
decomposed into the self-action and interaction parts. It is shown that the
interaction forces are attractive for both types of boundary conditions. At
separations between the plates larger than the curvature radius of the
background geometry they decay exponentially as functions of the proper
distance. The self-action force per unit surface of a single plate does not
depend on its location and depending on the boundary condition and on the
number of spatial dimensions can be either attractive or repulsive with
respect to the AdS boundary. By using the generalized zeta function
technique we also evaluate the total Casimir energy. Applications are given
in $Z_{2}$-symmetric braneworld models of the Randall-Sundrum type for
vector fields with even and odd parities.
\end{abstract}

\section{Introduction}

In a large number of physical problems the interactions of quantum fields
are expressed in terms of boundary conditions imposed on the field operator.
This idealization essentially simplifies the quantization and
renormalization procedures. For points outside of boundaries the structure
of divergences is the same as that in the boundary-free theory and
consequently the renormalization prescriptions for local physical
observables (like the expectation values of the energy-momentum tensor and
the current density) are the same as well. Additional divergences on
constraining boundaries are removed by renormalization of the physical
characteristics located on them. Depending on the specific problem, the
physical nature of boundaries can be different. The examples include
macroscopic bodies in quantum electrodynamics, interfaces separating
different phases of the theory, horizons in gravitational physics, branes in
higher dimensional models and so on. Another type of constraints on the
field operator, periodicity conditions, appear in field-theoretical models
with compact spatial dimensions. The boundary conditions on the field
operator modify the spectrum of the field fluctuations and as a consequence
the expectation values of physical quantities are shifted. This general
class of phenomena is known as the Casimir effect (for reviews see \cite%
{Most97}).

The physical characteristics in the Casimir effect depend on the field, on
the bulk and boundary geometries and on the boundary conditions imposed. In
particular, motivated by applications in gravity, cosmology and in condensed
matter systems, the investigations of the influence of the background
geometry are of special interest. Already in the case of free fields closed
analytic expressions for boundary-induced contributions in the expectation
values of physical observables are obtained for highly symmetric geometries.
Particularly, the de Sitter and anti-de Sitter (AdS) spacetimes have
attracted a great deal of attention. These spacetimes possess the same
number of symmetries as the Minkowski spacetime and, as a consequence, a
large number of physical problems are exactly solvable on their background.
This gives an idea on the influence of gravitational field on physical
phenomena in more complicated geometries.

In the present paper, as a background geometry we consider the AdS
spacetime. In addition to high symmetry, the importance of this geometry in
quantum field theory is motivated by a number of other reasons. The AdS
spacetime is not globally hyperbolic and the early interest was mainly
related to principal questions of the quantization procedure in such
spacetimes. To yield well defined dynamics, boundary conditions should be
chosen on timelike conformal infinity and this brings several qualitatively
new features compared to the Minkowskian theories. In particular, new type
of instabilities may arise. The importance of such studies is also due to
the appearance of AdS spacetime as a ground state in extended supergravity
and in string theories and as near horizon limit of extremal black holes and
black strings. The non-zero curvature of the AdS spacetime provides an
infrared regulator for correlation functions, consistent with supersymmetry
and modular invariance \cite{Call90}. Another new feature of the dynamics,
that distinguishes the AdS bulk from the Minkowski one, is the existence of
consistent theories for interacting higher spin fields. On top of all this,
the geometrical properties of the AdS spacetime play a crucial role in two
fascinating modern developments of high-energy physics. The first one, the
AdS/CFT\ correspondence (see \cite{Ahar00} for reviews), is a realization of
the holographic principle. It states a duality between theories formulated
in different numbers of spacetime dimensions: the supergravity or string
theory in AdS bulk and conformal field theory on its boundary. Among the
most important implications of this correspondence is the possibility for
the investigation of nonperturbative effects in one theory through the weak
coupling expansion of the dual theory. In addition to the high-energy
physics, the recent developments include applications in condensed matter
physics (holographic superconductors, quantum phase transitions, and
topological insulators) \cite{Pire14}. The second focus of intense interest
with AdS spacetime as the background geometry is various types of braneworld
models \cite{Maar10} where the standard model fields are restricted to a
hypersurface (brane) embedded in a higher dimensional spacetime. Initially
proposed for a resolution of the gauge hierarchy problem, these models give
new insights into various problems of particle physics and cosmology. The
existence of the branes on which the matter fields are confined is predicted
also by string theories.

The boundary-induced quantum vacuum effects for planar branes have been
widely studied in background of the AdS bulk for scalar \cite{Fabi00,Flac01}%
, fermionic \cite{Flac01b} and gauge \cite{Garr03,Teo10} fields. These
investigations were mainly motivated by the possibility of the radion field
stabilization in Randall-Sundrum type braneworld models by using the Casimir
forces acting on the branes, by generation of the cosmological constant on
the branes and by extensions of the AdS/CFT correspondence to the case with
boundaries in the conformal field theory side \cite{Nish09}. The vacuum
expectation values of the energy-momentum tensor for scalar and fermion
fields were investigated in \cite{Knap04}. The vacuum energy, the
energy-momentum tensor, and the current density for charged fields in
higher-dimensional models with compact subspaces have been considered in
\cite{Flac03,Beze15}. The models on the AdS bulk with de Sitter and AdS
branes have been discussed in \cite{Noji00}.

In the present paper we consider both the local and global effects for
quantum electromagnetic field induced by two parallel plates in the AdS bulk
with an arbitrary number of spatial dimensions. The propagators for vector
fields in AdS spacetime in the absence of additional boundaries/branes have
been considered in \cite{Alle86}. The different types of boundary conditions
for vector fields on the AdS boundary and their interpretations in the
context of AdS/CFT correspondence were considered in \cite{Ishi04}. The
dynamics of bulk gauge fields in the 5D Randall-Sundrum model have been
discussed in \cite{Davo00}. The electromagnetic Casimir energy and the
forces acting on the branes in 5D Randall-Sundrum model have been
investigated in \cite{Garr03,Teo10} for perfectly conducting boundary
condition. For the same boundary condition, the two-point functions and the
vacuum expectation value of the energy-momentum tensor in the geometry of a
single plate on AdS bulk with general number of spatial dimensions were
considered in \cite{Kota16}-\cite{Kota17b}. The electromagnetic Casimir
effect in de Sitter spacetime for planar boundaries has been discussed in
\cite{Saha14,Kota15}. The electromagnetic two-point functions and the
Casimir effect in background of Friedmann-Robertson-Walker cosmologies with
power-law scale factors were considered in \cite{Bell13}.

The outline of the paper is as follows. In the next section we describe the
geometry of the problem and present the mode functions. The two-point
functions for the vector potential and field strength tensor are presented
in section \ref{sec:2pVec} and \ref{sec:2pF}. By using the two-point
functions for the field strength tensor, the VEVs of the electric and
magnetic fields squared and the photon condensate are investigated in
section \ref{sec:2pE2}. The VEV of the energy-momentum tensor is discussed
in section \ref{sec:emt}. The Casimir forces acting on the plates are
considered in section \ref{sec:Force}. In section \ref{sec:VacEn} we
investigate the total vacuum energy by using the zeta function
regularization scheme. The applications of the obtained results to higher
dimensional generalizations of the Randall-Sundrum type braneworld models
are discussed in section \ref{sec:Branes}. The main results are summarized
in section \ref{sec:Conc}.

\section{Background geometry and the modes}

\label{sec:Modes}

We consider quantum electromagnetic field with the vector potential $A_{\mu
}(x)$, $\mu =0,1,\ldots ,D$, and with the field strength tensor $F_{\mu \nu
}=\partial _{\mu }A_{\nu }-\partial _{\nu }A_{\mu }$ in a $(D+1)$%
-dimensional spacetime with the metric tensor $g_{\mu \nu }$. The dynamics
of the field is governed by the Maxwell equation%
\begin{equation}
\nabla _{\nu }F^{\mu \nu }=\frac{1}{\sqrt{|g|}}\partial _{\nu }\left( \sqrt{%
|g|}F^{\mu \nu }\right) =0,  \label{Meq}
\end{equation}%
where $g=\mathrm{det}(g_{\mu \nu })$. We will assume that the vector
potential is constrained by the Lorentz condition
\begin{equation}
\nabla _{\mu }A^{\mu }=\frac{1}{\sqrt{|g|}}\partial _{\mu }\left( \sqrt{|g|}%
A^{\mu }\right) =0.  \label{gauge}
\end{equation}%
When boundaries are present, in order to have a well-posed Cauchy problem
one needs to impose appropriate boundary conditions on the field.

In the present paper the background geometry is the AdS spacetime generated
by a negative bulk cosmological constant $\Lambda $. In the Poincar\'{e}
coordinates $(x^{0}=t,x^{1},\ldots ,x^{D-1},x^{D}=z)$ the metric tensor is
given by
\begin{equation}
g_{\mu \nu }=\left( \frac{\alpha }{z}\right) ^{2}\eta _{\mu \nu },
\label{metric}
\end{equation}%
where $\eta _{\mu \nu }=\mathrm{diag}(1,-1,\ldots ,-1)$ is the metric tensor
for $(D+1)$-dimensional Minkowski spacetime. For the coordinates one has $%
-\infty <x^{i}<+\infty $ for $i=0,1,\ldots ,D-1$, and $0\leq z<\infty $. The
geometry is conformally related to the half of the Mnikowski spacetime. The
hypersurfaces $z=0$ and $z=\infty $ present the boundary and the horizon of
the AdS spacetime. The parameter $\alpha $ is expressed in terms of the
negative cosmological constant $\Lambda $ as $\alpha =\sqrt{(1-D)D/(2\Lambda
)}$. Instead of the coordinate $z$ one can introduce the coordinate $y$ in
accordance with $y=\alpha \ln (z/\alpha )$, $-\infty <y<+\infty $. In terms
of this coordinate the metric is given by $g_{\mu \nu }=\mathrm{diag}%
(e^{-2y/\alpha }\eta _{ik},-1)$ with $i,k=0,1,\ldots ,D-1$. The gauge
condition (\ref{gauge}) does not fix the vector potential uniquely. We will
impose an additional condition $A^{D}=0$ (for the consistency with (\ref%
{gauge}) see \cite{Teo10}). Under this condition the constraint (\ref{gauge}%
) is reduced to $\partial _{\mu }A^{\mu }=0$.

We are interested in the effects of two codimension one plates (branes in
braneworld models), parallel to the AdS horizon, on the properties of the
electromagnetic vacuum. The locations of the plates will be denoted by $%
z=z_{1}$ and $z=z_{2}$, $z_{2}>z_{1}$. If we denote by $y_{1}$ and $y_{2}$
the corresponding values of the $y$-coordinate, $y_{j}=\alpha \ln
(z_{j}/\alpha )$, $j=1,2$, then the proper distance between the plates is
expressed as $a=y_{2}-y_{1}=\alpha \ln (z_{2}/z_{1})$. Two types of boundary
conditions will be considered below. The first one is the higher-dimensional
generalization of the perfect conductor boundary condition in $3D$
electrodynamics and reads
\begin{equation}
n^{\mu _{1}}\,^{\ast }F_{\mu _{1}\cdots \mu _{D-1}}=0,\;z=z_{1},z_{2},
\label{BC}
\end{equation}%
where $^{\ast }F_{\mu _{1}\cdots \mu _{D-1}}=\varepsilon _{\mu \nu \mu
_{1}\cdots \mu _{D-1}}F^{\mu \nu }/(D-1)!$ is the dual of the field tensor
and $n^{\mu }$ is the normal vector to the boundary. As the second type of
boundary conditions we will take the condition
\begin{equation}
n^{\mu }F_{\mu \nu }=0,\;z=z_{1},z_{2}.  \label{BC2}
\end{equation}%
It is used in bag models of hadrons for the confinement of gluons inside the
bag. Both the conditions (\ref{BC}) and (\ref{BC2}) are gauge invariant.

In the problem under consideration, all the properties of the quantum vacuum
are encoded in two-point functions or the vacuum fluctuations correlators.
For the evaluation of the correlators we will use the mode-sum method. In
that method the two-point functions are presented in the form of a sum over
the products of the complete set of the electromagnetic modes obeying the
boundary conditions. We will denote the complete set for the vector
potential by $\{A_{(\beta )\mu },A_{(\beta )\mu }^{\ast }\}$, where $\beta $
corresponds to the set of quantum numbers specifying the modes and the star
stands for the complex conjugate. In accordance with the problem symmetry
the dependence of the modes on the coordinates $x^{l}$, $l=0,1,\ldots ,D-1$,
can be taken in the form $e^{ik_{l}x^{l}}$ with the wave vector components $%
k_{l}$. With this choice and by taking into account that $A_{(\beta )D}=0$,
the boundary condition (\ref{BC}) is reduced to $A_{(\beta )l}=0$ for $%
z=z_{1},z_{2}$, and from the condition (\ref{BC2}) we get $\partial
_{D}A_{(\beta )l}=0$, $z=z_{1},z_{2}$. This shows that in the geometry under
consideration the conditions (\ref{BC}) and (\ref{BC2}) are the analogs of
Dirichlet and Neumann boundary conditions for scalar fields. The $z$%
-dependence of the modes can be found from the field equation and the mode
functions for the vector potential are presented as
\begin{equation}
A_{(\beta )\mu }(x)=\epsilon _{(\sigma )\mu }z^{D/2-1}\left[
c_{1}J_{D/2-1}(\lambda z)+c_{2}Y_{D/2-1}(\lambda z)\right] e^{ik_{l}x^{l}},
\label{Modes}
\end{equation}%
where $J_{\nu }(x)$ and $Y_{\nu }(x)$ are the Bessel and Neumann functions, $%
k_{0}=\omega =\sqrt{\lambda ^{2}+k^{2}}$, $k^{2}=\sum_{l=1}^{D-1}k_{l}^{2}$.
The polarization vector $\epsilon _{(\sigma )\mu }$, with $\sigma =1,\ldots
,D-1$ corresponding to different polarizations, is normalized by the
condition $\eta ^{\mu \rho }\epsilon _{(\sigma )\mu }\epsilon _{(\sigma
^{\prime })\rho }=-\delta _{\sigma \sigma ^{\prime }}$. From the gauge
conditions it follows that $\epsilon _{(\sigma )D}=0$ and $\eta ^{\mu \rho
}k_{\mu }\epsilon _{(\sigma )\rho }=0$. The set $\beta $ of quantum numbers
is specified as $\beta =\sigma ,\lambda ,k_{1},\ldots ,k_{D-1}$. The
normalization condition for the modes (\ref{Modes}) has the form
\begin{equation}
\int d^{D}x\sqrt{|g|}[A_{(\beta ^{\prime })\mu }^{\ast }\nabla ^{0}A_{(\beta
)}^{\mu }-(\nabla ^{0}A_{(\beta ^{\prime })\mu }^{\ast })A_{(\beta )}^{\mu
}]=4i\pi \delta _{\beta \beta ^{\prime }},  \label{NC}
\end{equation}%
where $\delta _{\beta \beta ^{\prime }}$ is understood as the Kronecker
delta for discrete components of $\beta $ and the Dirac delta function for
the continuous ones. Note that for a scalar field with the curvature
coupling parameter $\xi $ and the mass $m$ the radial part of the mode
functions has the form $z^{D/2}\left[ c_{1}J_{\nu _{s}}(\lambda
z)+c_{2}Y_{\nu _{s}}(\lambda z)\right] $, where $\nu _{s}=\sqrt{%
D^{2}/4-D(D+1)\xi +m^{2}\alpha ^{2}}$. This shows that, unlike to the case
of the Minkowski bulk, for the AdS bulk the electromagnetic modes are not
reduced to the set of massless scalar modes with minimal or conformal
couplings.

The plates $z=z_{1}$ and $z=z_{2}$ divide the space into three parts: the
region between the AdS boundary and plate $z=z_{1}$, $0\leq z\leq z_{1}$
(region I), the region between the plates, $z_{1}\leq z\leq z_{2}$ (region
II), and the region between the plate $z=z_{2}$ and the horizon, $z_{2}\leq
z<\infty $ (region III). The coefficients $c_{1}$ and $c_{2}$ in (\ref{Modes}%
) depend on the region. We will consider the region between the plates. From
the boundary condition on $z=z_{1}$ it is seen that%
\begin{equation}
c_{1}=CY_{\nu }(\lambda z_{1}),\;c_{2}=-CJ_{\nu }(\lambda z_{1}),
\label{c12}
\end{equation}%
where%
\begin{equation}
\nu =\left\{
\begin{array}{cc}
D/2-1, & \text{for condition (\ref{BC}),} \\
D/2-2, & \text{for condition (\ref{BC2}),}%
\end{array}%
\right.  \label{nu}
\end{equation}%
and the constant $C$ is determined from the normalization condition (\ref{NC}%
). With the coefficients from (\ref{c12}), the mode functions in the region
II are expressed as%
\begin{equation}
A_{(\beta )\mu }=C\epsilon _{(\sigma )\mu }z^{D/2-1}g_{\nu ,D/2-1}(\lambda
z_{1},\lambda z)e^{ik_{l}x^{l}},  \label{Modes2}
\end{equation}%
where we have introduced the function%
\begin{equation}
g_{\nu ,\rho }(x,y)=Y_{\nu }(x)J_{\rho }(y)-J_{\nu }(x)Y_{\rho }(y).
\label{genu}
\end{equation}%
From the boundary condition on $z=z_{2}$ we find that the allowed values of $%
\lambda $ are roots of the equation%
\begin{equation}
g_{\nu ,\nu }(\lambda z_{1},\lambda z_{2})=0,  \label{lameig}
\end{equation}%
with $\nu $ from (\ref{nu}). The positive solutions of this equation with
respect to the first argument will be denoted by $\lambda _{\nu ,n}=\lambda
z_{1}$, $n=1,2,\ldots $, $\lambda _{\nu ,n+1}>\lambda _{\nu ,n}$. For the
eigenvalues of the energy one obtains%
\begin{equation}
\omega _{\nu ,n}=\sqrt{\lambda _{\nu ,n}^{2}/z_{1}^{2}+k^{2}}.  \label{omeq}
\end{equation}%
Note that for $\nu =\pm 1/2$ we have
\begin{equation}
g_{\pm 1/2,\pm 1/2}(\lambda z_{1},\lambda z)=-\frac{2\sin \left[ \lambda
(z-z_{1})\right] }{\pi \lambda \sqrt{zz_{1}}},  \label{g12}
\end{equation}%
and the corresponding eigenvalues are given by%
\begin{equation}
\lambda _{\pm 1/2,n}=\frac{\pi n}{z_{2}/z_{1}-1}.  \label{lamn}
\end{equation}%
These are the eigenvalues for the boundary condition (\ref{BC}) in the case $%
D=3$ and the eigenvalues for the boundary condition (\ref{BC2}) in $D=3,5$.
For the boundary condition (\ref{BC2}), the modes (\ref{lameig}) have been
discussed in \cite{Garr03,Teo10} within the framework of $D=4$
Randall-Sundrum setup.

By using the condition (\ref{lameig}), from (\ref{NC}), with the integration
over $z$ in the range $z\in \lbrack z_{1},z_{2}]$, for the coefficient $C$
one gets%
\begin{equation}
|C|^{2}=\frac{\lambda _{\nu ,n}T_{\nu }(\eta ,\lambda _{\nu ,n})}{8\left(
2\pi \right) ^{D-4}\alpha ^{D-3}z_{1}^{2}\omega _{\nu ,n}},  \label{C}
\end{equation}%
where
\begin{equation}
\eta =z_{2}/z_{1}=e^{a/\alpha },  \label{eta}
\end{equation}%
and we have introduced the notation%
\begin{equation}
T_{\nu }(\eta ,x)=x\left[ \frac{J_{\nu }^{2}(x)}{J_{\nu }^{2}(x\eta )}-1%
\right] ^{-1}.  \label{Te}
\end{equation}%
Note that $J_{\nu }(\lambda _{\nu ,n})/J_{\nu }(\lambda _{\nu ,n}\eta
)=Y_{\nu }(\lambda _{\nu ,n})/Y_{\nu }(\lambda _{\nu ,n}\eta )$ and the
expression (\ref{Te}) can also be written in terms of the Neumann function.
With the normalization constant (\ref{C}), the modes for the vector
potential are completely specified.

Let us consider the mode functions in the limit $z_{1}\rightarrow 0$ (the
left plate tends to the AdS boundary). For $\nu \geq 0$ one has $g_{\nu ,\nu
}(\lambda z_{1},\lambda z_{2})\approx Y_{\nu }(\lambda z_{1})J_{\nu
}(\lambda z_{2})$ and the equation determining the eigenvalues of the radial
quantum number $\lambda $ is reduced to $J_{\nu }(\lambda z_{2})=0$. By
taking into account (\ref{Te}), with the ratio of the Bessel functions
replace by the ratio of the Neumann functions, we can see that
\begin{equation}
Cg_{\nu ,D/2-1}(\lambda z_{1},\lambda z)\rightarrow C_{\mathrm{(I)}%
}J_{D/2-1}(\lambda z),  \label{ModesRegI}
\end{equation}%
with the normalization constant $|C_{\mathrm{(I)}}|^{2}=2/[\left( 2\pi
\right) ^{D-2}\alpha ^{D-3}\omega z_{2}^{2}J_{\nu }^{\prime 2}(\lambda
z_{2})]$. For the boundary condition (\ref{BC}) the mode functions obtained
in this way coincide with those considered in \cite{Saha16,Kota17b} for the
geometry of a single plate. For the boundary condition (\ref{BC2}) and for $%
D=3$ one has $\nu =-1/2$ the equation for the eigenvalues of $\lambda $ is
reduced to $Y_{\nu }(\lambda z_{2})=0$ which is equivalent to $\sin (\lambda
z_{2})=0$. For the radial part of the mode functions we get%
\begin{equation}
Cz^{D/2-1}g_{\nu ,D/2-1}(\lambda z_{1},\lambda z)\rightarrow \frac{\cos
(\lambda z)}{\sqrt{\pi z_{2}\omega }}.  \label{ModesRegIb}
\end{equation}%
Notice that for $D=3$ and in the case of the boundary condition (\ref{BC})
the eigenvalue equation is the same, $\sin (\lambda z_{2})=0$, and the mode
functions are obtained from (\ref{ModesRegIb}) by the replacement $\cos
(\lambda z)\rightarrow \sin (\lambda z)$.

In the geometry of a single plate at $z=z_{2}$ and for the region $0\leq
z\leq z_{2}$ the mode functions have the form (\ref{Modes}). The integral
over $z$ in the normalization condition (\ref{NC}) is reduced to $%
\int_{0}^{z_{2}}dz\,zZ_{D/2-1}(\lambda ^{\prime }z)Z_{D/2-1}(\lambda z)$,
where $Z_{\mu }(x)=c_{1}J_{\mu }(x)+c_{2}Y_{\mu }(x)$. For $c_{2}\neq 0$,
near the lower limit the integrand behaves as $z^{3-D}$. From here it
follows that for $D\geq 4$ and for normalizable modes one should take $%
c_{2}=0$. Imposing the boundary condition (\ref{BC}) or (\ref{BC2}) on $%
z=z_{2}$ we see that the eigenvalues for $\lambda $ are the roots of $J_{\nu
}(\lambda z_{2})=0$ with $\nu $ given by (\ref{nu}). These are the modes we
have obtained by the limiting transition $z_{1}\rightarrow 0$. For $D=3$ the
mode functions with $c_{2}\neq 0$ are normalizable and in order to uniquely
specify them an additional boundary condition is required on the AdS
boundary. As a result of the limiting transition $z_{1}\rightarrow 0$,
depending on the conditions imposed at $z=z_{1}$, we have obtained two
special types of boundary conditions.

\section{Two-point functions for the vector potential}

\label{sec:2pVec}

The two-point functions of a free field theory are important characteristics
of quantum fields describing the correlations of fluctuations at different
spacetime points. Having these functions one can evaluate the VEVs for
various physical quantities like the field squared, the energy-momentum
tensor and the current density. The free-field two-point functions are the
building blocks in the perturbative expansion of correlation functions in
interacting field theories. First we consider the two-point function for the
vector potential in the region between the plates (region II). With the mode
functions (\ref{Modes2}) and the eigenvalues of $\lambda $ determined from (%
\ref{lameig}), the two-point function (the positive-frequency Wightman
function)
\begin{equation}
\langle 0|A_{\mu }(x)A_{\rho }(x^{\prime })|0\rangle \equiv \langle A_{\mu
}A_{\rho }^{\prime }\rangle  \label{AAdef}
\end{equation}%
is evaluated by using the mode-sum formula%
\begin{equation}
\langle A_{\mu }A_{\rho }^{\prime }\rangle =\int d\mathbf{k}%
\,\sum_{n=1}^{\infty }\sum_{\sigma =1}^{D-1}A_{(\beta )\mu }(x)A_{(\beta
)\rho }(x^{\prime }),  \label{AA}
\end{equation}%
where $|0\rangle $ stands for the vacuum state and $\int d\mathbf{k}%
=\int_{-\infty }^{+\infty }dk_{1}\cdots \int_{-\infty }^{+\infty }dk_{D-1}$.
The summation over the polarizations is done by using the formula%
\begin{equation}
\sum_{\sigma =1}^{D-1}\epsilon _{(\sigma )\mu }\epsilon _{(\sigma )\rho }=%
\frac{k_{\mu }k_{\rho }}{\lambda ^{2}}-\eta _{\mu \rho },  \label{Polsum}
\end{equation}%
and for the nonzero components one finds
\begin{eqnarray}
\langle A_{\mu }A_{\rho }^{\prime }\rangle &=&\frac{\left( zz^{\prime
}\right) ^{D/2-1}}{8\left( 2\pi \right) ^{D-4}\alpha ^{D-3}}\int d\mathbf{k}%
\,\sum_{n=1}^{\infty }\frac{\lambda _{\nu ,n}T_{\nu }(\eta ,\lambda _{\nu
,n})}{\sqrt{\lambda _{\nu ,n}^{2}/z_{1}^{2}+k^{2}}}e^{ik_{l}\Delta x^{l}}
\notag \\
&&\times g_{\nu ,D/2-1}\left( \lambda _{\nu ,n},\lambda _{\nu
,n}z/z_{1}\right) g_{\nu ,D/2-1}\left( \lambda _{\nu ,n},\lambda _{\nu
,n}z^{\prime }/z_{1}\right) \left( \frac{k_{\mu }k_{\rho }}{\lambda _{\nu
,n}^{2}}-\frac{\eta _{\mu \rho }}{z_{1}^{2}}\right) .  \label{AA1}
\end{eqnarray}%
where $k_{0}=\omega _{\nu ,n}$, $\Delta x^{l}=x^{l}-x^{\prime l}$, $\mu
,\rho ,l=0,1,\ldots ,D-1$.

In (\ref{AA1}) the eigenvalues $\lambda _{\nu ,n}$ for general spatial
dimension are given implicitly and this representation is not convenient for
the evaluation of the VEVs in the coincidence limit of the arguments. A more
adapted representation is obtained by making use of a variant of the
generalized Abel-Plana formula \cite{Saha87,Saha08Book}%
\begin{equation}
\sum_{n=1}^{\infty }f(\lambda _{\nu ,n})T_{\nu }(\eta ,\lambda _{\nu ,n})=%
\frac{2}{\pi ^{2}}\int_{0}^{\infty }\frac{f(u)du}{J_{\nu }^{2}(u)+Y_{\nu
}^{2}(u)}-\frac{1}{2\pi }\int_{0}^{\infty }du\,\Omega _{\nu }^{(1)}(u,\eta u)%
\left[ f(iu)+f(-iu)\right] ,  \label{SumAbel}
\end{equation}%
with the notations
\begin{equation}
\Omega _{\nu }^{(1)}(x,y)=\frac{K_{\nu }(y)}{K_{\nu }(x)G_{\nu ,\nu }(x,y)},
\label{Om1}
\end{equation}%
and%
\begin{equation}
G_{\nu ,\rho }(x,y)=K_{\nu }(x)I_{\rho }(y)-(-1)^{\nu -\rho }I_{\nu
}(x)K_{\rho }(y).  \label{Gxy}
\end{equation}%
Here, $I_{\nu }(x)$ and $K_{\nu }(x)$ are the modified Bessel functions. In
the case of the function $f(u)$ corresponding to (\ref{AA1}), the conditions
of the validity for (\ref{SumAbel}) are satisfied if $z+z^{\prime }+|\Delta
t|<2z_{2}$. In the coincidence limit and for the region between the plates
this condition is obeyed for points away from the plate at $z=z_{2}$.

For the series in (\ref{AA1}) the function $f(u)$ is given by the expression%
\begin{equation}
f(u)=\frac{ue^{-i\sqrt{u^{2}/z_{1}^{2}+k^{2}}\Delta t}}{\sqrt{%
u^{2}/z_{1}^{2}+k^{2}}}g_{\nu ,D/2-1}\left( u,uz/z_{1}\right) g_{\nu
,D/2-1}\left( u,uz^{\prime }/z_{1}\right) \left( \frac{k_{\mu }k_{\rho }}{%
u^{2}}-\frac{\eta _{\mu \rho }}{z_{1}^{2}}\right) .  \label{fu}
\end{equation}%
By applying (\ref{SumAbel}), the two-point function is presented in the
decomposed form%
\begin{eqnarray}
\langle A_{\mu }A_{\rho }^{\prime }\rangle &=&\langle A_{\mu }A_{\rho
}^{\prime }\rangle _{1}+\frac{4\delta _{\nu }\left( zz^{\prime }\right)
^{D/2-1}}{\left( 2\pi \right) ^{D-1}\alpha ^{D-3}}\int d\mathbf{k}%
\,\int_{k}^{\infty }du\,  \notag \\
&&\times \left( \eta _{\mu \rho }u^{2}+\partial _{\mu }\partial _{\rho
}^{\prime }\right) e^{ik_{l}\Delta x^{l}}\,\frac{\Omega _{\nu
}^{(1)}(uz_{1},uz_{2})}{u\sqrt{u^{2}-k^{2}}}  \notag \\
&&\times G_{\nu ,D/2-1}\left( uz_{1},uz\right) G_{\nu ,D/2-1}\left(
uz_{1},uz^{\prime }\right) \cosh (\Delta t\sqrt{u^{2}-k^{2}}),  \label{AA2}
\end{eqnarray}%
for $\mu ,\rho =0,1,\ldots ,D-1$ and $l=1,\ldots ,D-1$ for the summation in $%
k_{l}\Delta x^{l}$. Here we have defined%
\begin{equation}
\delta _{\nu }=\left\{
\begin{array}{cc}
1, & \text{for }\nu =D/2-1, \\
-1, & \text{for }\nu =D/2-2.%
\end{array}%
\right.  \label{delnu}
\end{equation}%
The first term in the right-hand side of (\ref{AA2}) is given by the
expression%
\begin{eqnarray}
\langle A_{\mu }A_{\rho }^{\prime }\rangle _{1} &=&\frac{\left( zz^{\prime
}\right) ^{D/2-1}}{\left( 2\pi \right) ^{D-2}\alpha ^{D-3}}\int d\mathbf{k}%
\int_{0}^{\infty }d\lambda \,\frac{\lambda }{\omega }\left( \frac{k_{\mu
}k_{\rho }}{\lambda ^{2}}-\eta _{\mu \rho }\right)  \notag \\
&&\times \frac{g_{\nu ,D/2-1}\left( \lambda z_{1},\lambda z\right) g_{\nu
,D/2-1}\left( \lambda z_{1},\lambda z^{\prime }\right) }{J_{\nu
}^{2}(\lambda z_{1})+Y_{\nu }^{2}(\lambda z_{1})}e^{ik_{l}\Delta x^{l}},
\label{TwoAb}
\end{eqnarray}%
where $\omega =\sqrt{\lambda ^{2}+k^{2}}$ and $l=0,1,\ldots ,D-1$.

In the limit $z_{2}\rightarrow \infty $ the last term in (\ref{AA2}) tends
to zero and the part (\ref{TwoAb}) is interpreted as the two-point function
in the region $z>z_{1}$ for the geometry of a single plate at $z=z_{1}$.
This can also be seen directly by evaluating the corresponding mode-sum. The
mode functions are still given by (\ref{Modes2}) where now the eigenvalues
of $\lambda $ are continuous. The normalization constant is obtained from (%
\ref{NC}) with the $z$-integral over the region $z_{1}\leq z<\infty $ and is
given by
\begin{equation}
|C_{\mathrm{(III)}}|^{2}=\frac{(2\pi )^{2-D}\alpha ^{3-D}\lambda }{\omega %
\left[ J_{\nu }^{2}(\lambda z_{1})+Y_{\nu }^{2}(\lambda z_{1})\right] }.
\label{CIII}
\end{equation}%
The mode-sum with the functions (\ref{Modes2}) and the normalization
coefficient (\ref{CIII}) leads to the representation (\ref{TwoAb}). The
boundary-induced contribution in (\ref{TwoAb}) can be separated in a way
similar to that used in \cite{Kota17} for the special case of the boundary
condition (\ref{BC}). The corresponding expression reads%
\begin{eqnarray}
\langle A_{\mu }A_{\rho }^{\prime }\rangle _{1} &=&\langle A_{\mu }A_{\rho
}^{\prime }\rangle _{0}+\frac{4\delta _{\nu }\left( zz^{\prime }\right)
^{D/2-1}}{\left( 2\pi \right) ^{D-1}\alpha ^{D-3}}\int d\mathbf{k}%
\,\int_{k}^{\infty }du\,\left( \eta _{\mu \rho }x^{2}+\partial _{\mu
}\partial _{\rho }^{\prime }\right) e^{ik_{l}\Delta x^{l}}  \notag \\
&&\times \frac{\cosh (\Delta t\sqrt{u^{2}-k^{2}})}{u\sqrt{u^{2}-k^{2}}}\frac{%
I_{\nu }(uz_{1})}{K_{\nu }(uz_{1})}K_{D/2-1}(uz)K_{D/2-1}(uz^{\prime }),
\label{AA11}
\end{eqnarray}%
where $l=0,1,\ldots ,D-1$, and $\langle A_{\mu }A_{\rho }^{\prime }\rangle
_{0}$ is the two-point function in AdS spacetime when the plates are absent.

An alternative representation for the two-point function of the vector
potential in the region between the plates is obtained by using the relation%
\begin{eqnarray}
&&\frac{I_{\nu }(uz_{1})}{K_{\nu }(uz_{1})}K_{D/2-1}(uz)K_{D/2-1}(uz^{\prime
})-\frac{K_{\nu }(uz_{2})}{I_{\nu }(uz_{2})}I_{D/2-1}(uz)I_{D/2-1}(uz^{%
\prime })  \notag \\
&& =\sum_{j=1,2}(-1)^{j}\Omega _{\nu }^{(j)}(uz_{1},uz_{2})G_{\nu
,D/2-1}\left( uz_{j},uz\right) G_{\nu ,D/2-1}\left( uz_{j},uz^{\prime
}\right) ,  \label{Rel1}
\end{eqnarray}%
where%
\begin{equation}
\Omega _{\nu }^{(2)}(x,y)=\frac{I_{\nu }(x)}{I_{\nu }(y)G_{\nu ,\nu }(x,y)}.
\label{Om2}
\end{equation}%
This leads to the expression%
\begin{eqnarray}
\langle A_{\mu }A_{\rho }^{\prime }\rangle &=&\langle A_{\mu }A_{\rho
}^{\prime }\rangle _{2}+\frac{4\delta _{\nu }\left( zz^{\prime }\right)
^{D/2-1}}{\left( 2\pi \right) ^{D-1}\alpha ^{D-3}}\int d\mathbf{k}%
\,\int_{k}^{\infty }du\,  \notag \\
&&\times \left( \eta _{\mu \rho }u^{2}+\partial _{\mu }\partial _{\rho
}^{\prime }\right) e^{ik_{l}\Delta x^{l}}\,\frac{\Omega _{\nu
}^{(2)}(uz_{1},uz_{2})}{u\sqrt{u^{2}-k^{2}}}  \notag \\
&&\times G_{\nu ,D/2-1}\left( uz_{2},uz\right) G_{\nu ,D/2-1}\left(
uz_{2},uz^{\prime }\right) \cosh (\Delta t\sqrt{u^{2}-k^{2}}),  \label{AA3}
\end{eqnarray}%
where%
\begin{eqnarray}
\langle A_{\mu }A_{\rho }^{\prime }\rangle _{2} &=&\langle A_{\mu }A_{\rho
}^{\prime }\rangle _{0}+\frac{4\delta _{\nu }\left( zz^{\prime }\right)
^{D/2-1}}{\left( 2\pi \right) ^{D-1}\alpha ^{D-3}}\int d\mathbf{k}%
\,\int_{k}^{\infty }du\,\left( \eta _{\mu \rho }u^{2}+\partial _{\mu
}\partial _{\rho }^{\prime }\right) e^{ik_{l}\Delta x^{l}}  \notag \\
&&\times \frac{\cosh (\Delta t\sqrt{u^{2}-k^{2}})}{u\sqrt{u^{2}-k^{2}}}\frac{%
K_{\nu }(uz_{2})}{I_{\nu }(uz_{2})}I_{D/2-1}(uz)I_{D/2-1}(uz^{\prime }),
\label{AA21}
\end{eqnarray}%
is the two-point function in the region $0\leqslant z\leqslant z_{2}$ for
the geometry of a single plate at $z=z_{2}$.

In the region I the two-point function is given by (\ref{AA21}) with the
replacement $z_{2}\rightarrow z_{1}$. For the region III, $z_{2}\leqslant
z<\infty $, the two-point function is obtained from (\ref{AA11}) replacing $%
z_{2}\rightarrow z_{1}$. The corresponding result could also be obtained
directly considering the field dynamics in that region. The mode functions
are given by (\ref{Modes2}) with the replacement (\ref{ModesRegI}) and $%
z_{2}\rightarrow z_{1}$ and the eigenvalues for $\lambda $ are the zeros of
the function $J_{\nu }(\lambda z_{1})$. The series over these zeros in the
mode sum for the two-point function $\langle A_{\mu }A_{\rho }^{\prime
}\rangle _{1}$ is summed by using the respective formula from \cite%
{Saha87,Saha08Book}. That allows to separate the plate-induced contribution
explicitly.

\section{Correlators for the field tensor}

\label{sec:2pF}

The VEVs of physical observables bilinear in the fields, such as the fields
squared and energy-momentum tensor, are obtained from the two-point function
for the field strength tensor%
\begin{equation}
\langle 0|F_{\mu \sigma }(x)F_{\kappa \rho }(x^{\prime })|0\rangle =\langle
F_{\mu \sigma }F_{\kappa \rho }^{\prime }\rangle .  \label{FFd}
\end{equation}%
Given the plate-induced contributions to the two-point function for the
vector potential we can find the corresponding contributions to the
correlator (\ref{FFd}) by differentiations. First of all for the components
with $\mu ,\sigma ,\kappa ,\rho =1,...,D-1$ we get%
\begin{eqnarray}
\langle F_{\mu \sigma }F_{\kappa \rho }^{\prime }\rangle &=&\langle F_{\mu
\sigma }F_{\kappa \rho }^{\prime }\rangle _{j}+\frac{8\delta _{\nu }\left(
zz^{\prime }\right) ^{D/2-1}}{\left( 2\pi \right) ^{D-1}\alpha ^{D-3}}\int d%
\mathbf{k}\,e^{ik_{l}\Delta x^{l}}\int_{k}^{\infty }du\,u  \notag \\
&&\times \frac{\cosh (\Delta t\sqrt{u^{2}-k^{2}})}{\sqrt{u^{2}-k^{2}}}F_{\nu
}^{(j)}(uz,uz^{\prime })\left( \eta _{\lbrack \sigma \rho }k_{\mu
]}k_{\kappa }+\eta _{\lbrack \mu \kappa }k_{\sigma ]}k_{\rho }\right) ,
\label{FF1}
\end{eqnarray}%
\begin{eqnarray}
\langle F_{\mu D}F_{\kappa D}^{\prime }\rangle &=&\langle F_{\mu D}F_{\kappa
D}^{\prime }\rangle _{j}+\frac{4\delta _{\nu }\alpha ^{3-D}}{\left( 2\pi
\right) ^{D-1}}\int d\mathbf{k}\,e^{ik_{l}\Delta x^{l}}\int_{k}^{\infty
}du\,u\left( \frac{k_{\mu }k_{\kappa }}{u^{2}}+\eta _{\mu \kappa }\right)
\notag \\
&&\times \frac{\cosh (\Delta t\sqrt{u^{2}-k^{2}})}{\sqrt{u^{2}-k^{2}}}%
\partial _{z}\partial _{z^{\prime }}\left[ (zz^{\prime })^{D/2-1}F_{\nu
}^{(j)}(uz,uz^{\prime })\right] ,  \label{FFDD}
\end{eqnarray}%
where $j=1,2$, $l=1,\ldots ,D-1$, the square brackets in the index
expression mean anti-symmetrization with respect to the corresponding
indices (indices $\sigma $ and $\mu $ in (\ref{FF1})) and%
\begin{equation}
F_{\nu }^{(j)}(xz,xz^{\prime })=\Omega _{\nu }^{(j)}(xz_{1},xz_{2})G_{\nu
,D/2-1}\left( xz_{j},xz\right) G_{\nu ,D/2-1}\left( xz_{j},xz^{\prime
}\right) .  \label{Fj}
\end{equation}%
Here, $j=1$ and $j=2$ provide two equivalent representations. These
representations are also valid for $\mu =\kappa =0$ if $k_{0}$ is understood
as $k_{0}=\sqrt{k^{2}-u^{2}}$. For the components with only one of the
indices being $0$ and $D$ we have%
\begin{eqnarray}
\langle F_{0\mu }F_{\rho \kappa }^{\prime }\rangle &=&\langle F_{0\mu
}F_{\rho \kappa }^{\prime }\rangle _{j}-\frac{8i\delta _{\nu }\left(
zz^{\prime }\right) ^{D/2-1}}{\left( 2\pi \right) ^{D-1}\alpha ^{D-3}}\int d%
\mathbf{k}\,e^{ik_{l}\Delta x^{l}}  \notag \\
&&\times \int_{k}^{\infty }du\,u\sinh (\Delta t\sqrt{u^{2}-k^{2}})F_{\nu
}^{(j)}(uz,uz^{\prime })\delta _{\mu \lbrack \kappa }k_{\rho ]},
\label{FF0nu}
\end{eqnarray}%
\begin{eqnarray}
\langle F_{D\mu }F_{\rho \kappa }^{\prime }\rangle &=&\langle F_{D\mu
}F_{\rho \kappa }^{\prime }\rangle _{j}-\frac{8i\delta _{\nu }z^{\prime
D/2-1}}{\left( 2\pi \right) ^{D-1}\alpha ^{D-3}}\int d\mathbf{k}%
\,e^{ik_{l}\Delta x^{l}}\int_{k}^{\infty }du\,u  \notag \\
&&\times \frac{\cosh (\sqrt{u^{2}-k^{2}}\Delta t)}{\sqrt{u^{2}-k^{2}}}%
\partial _{z}\left[ z^{D/2-1}F_{\nu }^{(j)}(uz,uz^{\prime })\right] \delta
_{\mu \lbrack \kappa }k_{\rho ]}.  \label{FFDnu}
\end{eqnarray}%
The remaining components are found by using the relation%
\begin{equation}
\langle F_{\mu \sigma }F_{\kappa \rho }^{\prime }\rangle =\langle F_{\kappa
\rho }^{\prime }F_{\mu \sigma }\rangle ^{\ast },  \label{FFrel}
\end{equation}%
valid for all values of the indices. The parts $\langle F_{\mu \sigma
}F_{\kappa \rho }^{\prime }\rangle _{j}$ in (\ref{FF1})-(\ref{FFDnu})
present the two-point functions in the geometry for a single plate at $%
z=z_{j}$. They are decomposed as
\begin{equation}
\langle F_{\mu \sigma }F_{\kappa \rho }^{\prime }\rangle _{j}=\langle F_{\mu
\sigma }F_{\kappa \rho }^{\prime }\rangle _{0}+\langle F_{\mu \sigma
}F_{\kappa \rho }^{\prime }\rangle _{j}^{\mathrm{(b)}},  \label{FFj}
\end{equation}%
where $\langle F_{\mu \sigma }F_{\kappa \rho }^{\prime }\rangle _{0}$ is the
corresponding function in the geometry without plates and the part $\langle
F_{\mu \sigma }F_{\kappa \rho }^{\prime }\rangle _{j}^{\mathrm{(b)}}$ is
induced by a single plate at $z=z_{j}$. The expressions for $\langle F_{\mu
\sigma }F_{\kappa \rho }^{\prime }\rangle _{j}^{\mathrm{(b)}}$ are obtained
from the last terms in (\ref{FF1}), (\ref{FFDD}), (\ref{FF0nu}) and (\ref%
{FFDnu}) by the replacement%
\begin{equation}
F_{\nu }^{(j)}(xz,xz^{\prime })\rightarrow \frac{K_{\nu }(uz_{j})}{I_{\nu
}(uz_{j})}I_{D/2-1}(uz)I_{D/2-1}(uz^{\prime }),  \label{Repl1}
\end{equation}%
for the region $z,z^{\prime }<z_{j}$ and by the replacement%
\begin{equation}
F_{\nu }^{(j)}(xz,xz^{\prime })\rightarrow \frac{I_{\nu }(uz_{j})}{K_{\nu
}(uz_{j})}K_{D/2-1}(uz)K_{D/2-1}(uz^{\prime }),  \label{Repl2}
\end{equation}%
in the region $z,z^{\prime }>z_{j}$. The last terms in (\ref{FF1}), (\ref%
{FFDD}), (\ref{FF0nu}) and (\ref{FFDnu}) are induced by the plate $%
z=z_{j^{\prime }}$, $j^{\prime }=1,2$, $j^{\prime }\neq j$, when one adds it
to the geometry with a single plate at $z=z_{j}$.

In the evaluation of the local VEVs we need the two-point functions in the
coincidence limit $x^{\prime }\rightarrow x$. For $\mu ,\sigma ,\kappa ,\rho
=0,1,\ldots ,D-1$ and in the coincidence limit, for the nonzero components
one finds%
\begin{eqnarray}
\langle F_{\mu \sigma }F_{\kappa \rho }\rangle &=&\langle F_{\mu \sigma
}F_{\kappa \rho }\rangle _{j}-\frac{2\delta _{\nu }\left( \eta _{\mu \kappa
}\eta _{\sigma \rho }-\eta _{\mu \rho }\eta _{\sigma \kappa }\right) z^{D-2}%
}{\left( 4\pi \right) ^{D/2-1}\Gamma (D/2+1)\alpha ^{D-3}}  \notag \\
&&\times \int_{0}^{\infty }du\,u^{D+1}\Omega _{\nu
}^{(j)}(uz_{1},uz_{2})G_{\nu ,D/2-1}^{2}\left( uz_{j},uz\right) ,
\label{FFc1}
\end{eqnarray}%
\begin{eqnarray}
\langle F_{\mu D}F_{\kappa D}\rangle &=&\langle F_{\mu D}F_{\kappa D}\rangle
_{j}+\frac{\delta _{\nu }\eta _{\mu \kappa }(D-1)\alpha ^{3-D}z^{D-2}}{%
\left( 4\pi \right) ^{D/2-1}\Gamma (D/2+1)}  \notag \\
&&\times \int_{0}^{\infty }du\,u^{D+1}\Omega _{\nu
}^{(j)}(uz_{1},uz_{2})G_{\nu ,D/2-2}^{2}\left( uz_{j},uz\right) ,
\label{FFc2}
\end{eqnarray}%
where the relation
\begin{equation}
\partial _{z}\left[ z^{D/2-1}G_{\nu ,D/2-1}(uz_{j},uz)\right]
=uz^{D/2-1}G_{\nu ,D/2-2}\left( uz_{j},uz\right) ,  \label{RelG12}
\end{equation}%
has been used. The boundary-induced contributions in the VEVs $\langle
F_{\mu \sigma }F_{\kappa \rho }\rangle _{j}$ for the geometry of a single
plate at $z=z_{j}$ are obtained from the last terms in (\ref{FFc1}) and (\ref%
{FFc2}) by the replacements (\ref{Repl1}) and (\ref{Repl2}) with $z^{\prime
}=z$. Note that for points away from the boundaries the divergences in the
coincidence limit are contained in the parts $\langle F_{\mu \sigma
}F_{\kappa \rho }\rangle _{0}$ only. In the representations given above the
contribution $\langle F_{\mu \sigma }F_{\kappa \rho }\rangle _{0}$ is
explicitly extracted and consequently the renormalization of the local
observables at points outside of plates is reduced to the renormalization in
the geometry without plates. In the following sections the expressions (\ref%
{FFc1}) and (\ref{FFc2}) are used for the evaluation of the VEVs of fields
squared and of the energy-momentum tensor.

\section{VEVs of the electric and magnetic fields squared and photon
condensate}

\label{sec:2pE2}

We start the investigation of the local VEVs from the electric field
squared. Having the two-point functions of the field tensor in the
coincidence limit, this VEV is evaluated by using the relation
\begin{equation}
\langle E^{2}\rangle =-g^{00}g^{\mu \rho }\langle F_{0\mu }F_{0\rho }\rangle
.  \label{E2}
\end{equation}%
From (\ref{FFc1}) and (\ref{FFc2}), in the region $z_{1}<z<z_{2}$ one gets%
\begin{eqnarray}
\langle E^{2}\rangle &=&\langle E^{2}\rangle _{j}+\frac{\delta _{\nu
}(D-1)\alpha ^{-1-D}z^{D+2}}{\left( 4\pi \right) ^{D/2-1}\Gamma (D/2+1)}%
\int_{0}^{\infty }dx\,x^{D+1}\Omega _{\nu }^{(j)}(xz_{1},xz_{2})  \notag \\
&&\times \left[ 2G_{\nu ,D/2-1}^{2}\left( xz_{j},xz\right) +G_{\nu
,D/2-2}^{2}\left( xz_{j},xz\right) \right] .  \label{E21}
\end{eqnarray}%
Here, $\langle E^{2}\rangle _{j}$ is the VEV of the electric field squared
for the geometry of a single plate at $z=z_{j}$ in the region $z>z_{1}$ for $%
j=1$ and $z<z_{2}$ for $j=2$. For $j=1$ the VEV is given by the expressions%
\begin{equation}
\langle E^{2}\rangle _{1}=\langle E^{2}\rangle _{0}+\frac{\delta _{\nu
}\left( D-1\right) \alpha ^{-1-D}z^{D+2}}{\left( 4\pi \right) ^{D/2-1}\Gamma
(D/2+1)}\int_{0}^{\infty }dx\,x^{D+1}\frac{I_{\nu }(xz_{1})}{K_{\nu }(xz_{1})%
}\left[ 2K_{D/2-1}^{2}(xz)+K_{D/2-2}^{2}(xz)\right] ,  \label{E211}
\end{equation}%
where $\langle E^{2}\rangle _{0}$ is the VEV\ in the absence of the plates.
For $j=2$ the expression $\langle E^{2}\rangle _{2}$ in (\ref{E21}) is
obtained from (\ref{E211}) by the replacements $z_{1}\rightarrow z_{2}$ and $%
I\rightleftarrows K$. In the region $z<z_{1}$ the VEV\ of the electric field
squared is given by (\ref{E211}) with the replacements $I\rightleftarrows K$
(in the special case of boundary condition (\ref{BC}) the sign difference
with the result in \cite{Saha16} is related to the definition of $\langle
E^{2}\rangle $ in (\ref{E2}) without the minus sign in the right-hand side).
The VEV for the region $z>z_{2}$ is obtained from (\ref{E211}) by the
replacement $z_{1}\rightarrow z_{2}$. The second term in the right-hand side
of (\ref{E21}) is interpreted as the contribution induced by the second
plate when we add it to the geometry of a single plate at $z=z_{j}$. Note
that $G_{\mu ,\mu }(x,y)>0$ for $y>x>0$ and, hence, $\Omega _{\nu
}^{(j)}(x,y)>0$. From here we conclude that both the single plate induced
(the last term in (\ref{E211})) and the second plate induced (the last term
in (\ref{E21})) contributions to the VEV of the electric field squared are
positive/negative for the boundary condition (\ref{BC})/(\ref{BC2}).

The VEV (\ref{E21}) diverges on the plates. The divergences come from the
single plate contributions $\langle E^{2}\rangle _{j}$ and the second plate
contribution in (\ref{E21}) is finite on the plate $z=z_{j}$. In order to
find the leading term in the asymptotic expansion over the distance from the
plate we note that for points near the plate the dominant contribution to
the integral in (\ref{E211}) (and in the analog integral for $\langle
E^{2}\rangle _{2}$) comes from large values of $x$ and we can use the
corresponding asymptotics for the modified Bessel functions. For $%
|z/z_{j}-1|\ll 1$, to the leading order this gives%
\begin{equation}
\langle E^{2}\rangle \approx \langle E^{2}\rangle _{j}\approx \delta _{\nu }%
\frac{3\left( D-1\right) \Gamma ((D+1)/2)\alpha ^{-1-D}}{2\left( 4\pi
\right) ^{(D-1)/2}|1-z_{j}/z|^{D+1}}.  \label{Enear}
\end{equation}%
Note that in the asymptotic region under consideration $|1-z_{j}/z|\approx
|y-y_{j}|/\alpha $ and in terms of the coordinate $y$ the leading term does
not depend on the curvature radius. To the leading order the VEV coincides
with that for plates in the Minkowski spacetime (see below). This feature
was expected by taking into account that near the plates the dominant
contribution to the VEVs comes from the fluctuations with small wavelengths
(compared with the curvature radius) and the influence of the gravitational
field on those modes is weak.

In a similar way we can find the VEV\ $\langle F_{\mu \sigma }F^{\mu \sigma
}\rangle $. It is the analog of the gluon condensate in quantum
chromodynamics and is known as photon condensate (see, for instance, \cite%
{Vain89}). From (\ref{FFc1}) and (\ref{FFc2}) we find%
\begin{eqnarray}
\left\langle F_{\mu \sigma }F^{\mu \sigma }\right\rangle &=&\left\langle
F_{\mu \sigma }F^{\mu \sigma }\right\rangle _{j}-\frac{4\delta _{\nu }\left(
D-1\right) z^{D+2}}{\left( 4\pi \right) ^{D/2-1}\Gamma (D/2)\alpha ^{D+1}}%
\int_{0}^{\infty }dx\,x^{D+1}\,  \notag \\
&&\times \Omega _{\nu }^{(j)}(xz_{1},xz_{2})\left[ G_{\nu ,D/2-1}^{2}\left(
uz_{j},uz\right) +G_{\nu ,D/2-2}^{2}\left( uz_{j},uz\right) \right] ,
\label{F2}
\end{eqnarray}%
where%
\begin{eqnarray}
\left\langle F_{\mu \sigma }F^{\mu \sigma }\right\rangle _{1}
&=&\left\langle F_{\mu \sigma }F^{\mu \sigma }\right\rangle _{0}-\frac{%
4\delta _{\nu }\left( D-1\right) z^{D+2}}{\left( 4\pi \right) ^{D/2-1}\Gamma
(D/2)\alpha ^{D+1}}\int_{0}^{\infty }dx\,x^{D+1}  \notag \\
&&\times \frac{I_{\nu }(xz_{1})}{K_{\nu }(xz_{1})}\left[
K_{D/2-1}^{2}(xz)+K_{D/2-2}^{2}(xz)\right] ,  \label{F21}
\end{eqnarray}%
is the corresponding VEV in the region $z>z_{1}$ for the geometry of a
single plate at $z=z_{1}$. For $j=2$, the term $\left\langle F_{\beta \sigma
}F^{\beta \sigma }\right\rangle _{2}$ is the VEV in the geometry of a single
plate at $z=z_{2}$ for the region $z<z_{2}$. The boundary induced
contribution to the photon condensate is negative for the boundary condition
(\ref{BC}) and positive for the condition (\ref{BC2}). Near the plates one
has the asymptotic $\left\langle F_{\mu \sigma }F^{\mu \sigma }\right\rangle
\approx -2\left\langle E^{2}\right\rangle /3$ with $\left\langle
E^{2}\right\rangle $ given by (\ref{Enear}).

Having the VEVs of the electric field squared and the photon condensate we
can find the VEV of the magnetic field squared by using the relation $%
\langle B^{2}\rangle =\langle E^{2}\rangle +\langle F_{\mu \sigma }F^{\mu
\sigma }\rangle /2$. Note that in dimensions $D>3$ the magnetic field is not
a spatial vector and the corresponding VEV is given by%
\begin{equation}
\langle B^{2}\rangle =\frac{1}{2}g^{lm}g^{np}\langle F_{ln}F_{mp}\rangle ,
\label{B2}
\end{equation}%
with $l,m,n,p=1,2,\ldots ,D$. From (\ref{E21}) and (\ref{F2}) one gets%
\begin{eqnarray}
\langle B^{2}\rangle &=&\langle B^{2}\rangle _{j}-\frac{\delta _{\nu
}(D-1)\alpha ^{-1-D}z^{D+2}}{\left( 4\pi \right) ^{D/2-1}\Gamma (D/2+1)}%
\int_{0}^{\infty }dx\,x^{D+1}\Omega _{\nu }^{(j)}(xz_{1},xz_{2})  \notag \\
&&\times \left[ \left( D-2\right) G_{\nu ,D/2-1}^{2}\left( xz_{j},xz\right)
+\left( D-1\right) G_{\nu ,D/2-2}^{2}\left( xz_{j},xz\right) \right] .
\label{B21}
\end{eqnarray}%
Here the part corresponding to a single plate geometry is defined as%
\begin{eqnarray}
\langle B^{2}\rangle _{1} &=&\langle B^{2}\rangle _{0}-\frac{\delta _{\nu
}\left( D-1\right) \alpha ^{-1-D}z^{D+2}}{\left( 4\pi \right) ^{D/2-1}\Gamma
(D/2+1)}\int_{0}^{\infty }dx\,x^{D+1}\frac{I_{\nu }(xz_{1})}{K_{\nu }(xz_{1})%
}  \notag \\
&&\times \left[ (D-2)K_{D/2-1}^{2}(xz)+(D-1)K_{D/2-2}^{2}(xz)\right] ,
\label{B22}
\end{eqnarray}%
for $j=1$ and the expression for $\langle B^{2}\rangle _{2}$ is obtained
from (\ref{B22}) by the replacements $I\rightleftarrows K$ and $%
z_{1}\rightarrow z_{2}$ in the boundary-induced contribution. In a way
similar to that for the electric field squared we can see that the single
plate induced (the last term in (\ref{B22})) and the second plate induced
(the last term in (\ref{B21})) contributions to the VEV\ of the magnetic
field squared are negative/positive for the boundary condition (\ref{BC})/(%
\ref{BC2}). Near the plates the VEV of the magnetic field squared behaves
like $\langle B^{2}\rangle \approx (1-2D/3)\langle E^{2}\rangle $, where the
asymptotic for $\langle E^{2}\rangle $ near the plate $z=z_{j}$ is given by (%
\ref{Enear}).

For $D=3$, by using the expressions for the functions $I_{\pm 1/2}(x)$, $%
K_{\pm 1/2}(x)$, we can see that
\begin{eqnarray}
G_{\pm 1/2,\pm 1/2}(xz_{j},xz) &=&\frac{\sinh \left[ x\left( z-z_{j}\right) %
\right] }{x\sqrt{zz_{j}}},  \notag \\
G_{\pm 1/2,\mp 1/2}(xz_{j},xz) &=&\frac{\cosh \left[ x\left( z-z_{j}\right) %
\right] }{x\sqrt{zz_{j}}}.  \label{G12}
\end{eqnarray}%
With these expressions, from (\ref{E21}), (\ref{E211}), (\ref{B21}) and (\ref%
{B22}) one finds%
\begin{equation}
\langle F^{2}\rangle =\langle F^{2}\rangle _{0}-\frac{3\delta _{(F)}}{4\pi
\alpha ^{4}}-\frac{z^{4}\delta _{(F)}}{4\pi \alpha ^{4}}\left[ \frac{\delta
_{(F)}\pi ^{4}/45}{\left( z_{2}-z_{1}\right) ^{4}}\mp \frac{3}{\left(
z-z_{1}\right) ^{4}}\mp \int_{0}^{\infty }dx\,\frac{x^{3}\cosh \left[
x\left( z-z_{1}\right) \right] }{e^{x\left( z_{2}-z_{1}\right) }-1}\right] ,
\label{F2D3}
\end{equation}%
where $F=E,B$, for the electric and magnetic fields respectively and%
\begin{equation}
\delta _{(E)}=-\delta _{(B)}=1.  \label{delEB}
\end{equation}%
The upper and lower signs in (\ref{F2D3}) correspond to the boundary
conditions (\ref{BC}) and (\ref{BC2}), respectively. An alternative
expression is obtained by using the expansion $1/(e^{u}-1)=\sum_{n=1}^{%
\infty }e^{-nu}$ in the integral term:%
\begin{equation}
\langle F^{2}\rangle =\langle F^{2}\rangle _{0}-\frac{3\delta _{(F)}}{4\pi
\alpha ^{4}}-\frac{\alpha ^{-4}z^{4}}{4\pi \left( z_{2}-z_{1}\right) ^{4}}%
\left[ \frac{\pi ^{4}}{45}\mp 3\delta _{(F)}\sum_{n=-\infty }^{\infty
}\left( n-\frac{z-z_{1}}{z_{2}-z_{1}}\right) ^{-4}\right] ,  \label{F2D3b}
\end{equation}%
again, with the upper and lower signs corresponding to the conditions (\ref%
{BC}) and (\ref{BC2}).

In the Minkowskian limit, corresponding to $\alpha \rightarrow \infty $ for
fixed $y$, one has $z\approx \alpha +y$ and $z$, $z_{j}$ are large. By using
the asymptotic expressions for the modified Bessel functions for large
values of the argument, we can see that%
\begin{eqnarray}
G_{\mu ,\mu }(xz_{1},xz) &\approx &\frac{\sinh \left[ x\left( y-y_{1}\right) %
\right] }{\alpha x},  \notag \\
G_{\mu ,\mu \pm 1}\left( xz_{1},xz\right) &\approx &\frac{\cosh \left[
x\left( y-y_{1}\right) \right] }{\alpha x},  \label{Gmink}
\end{eqnarray}%
and
\begin{equation}
\Omega _{\nu }^{(j)}(xz_{1},xz_{2})\approx \frac{2\alpha x}{e^{2ax}-1}.
\label{OmMink}
\end{equation}%
Substituting these expressions in (\ref{E21}) and (\ref{B21}), to the
leading order one gets $\langle F^{2}\rangle \rightarrow \langle
F^{2}\rangle _{\mathrm{M}}$, $F=E,B$, with%
\begin{eqnarray}
\langle E^{2}\rangle _{\mathrm{M}} &=&-\frac{(D-1)\Gamma \left( \left(
D+1\right) /2\right) }{\left( 4\pi \right) ^{(D-1)/2}a^{D+1}}\left[ \zeta
(D+1)\mp \sum_{n=-\infty }^{\infty }\frac{3/2}{\left\vert
n-(y-y_{1})/a\right\vert ^{D+1}}\right] ,  \notag \\
\langle B^{2}\rangle _{\mathrm{M}} &=&-\frac{(D-1)\Gamma \left( \left(
D+1\right) /2\right) }{\left( 4\pi \right) ^{(D-1)/2}a^{D+1}}\left[ \zeta
(D+1)\pm \sum_{n=-\infty }^{\infty }\frac{D-3/2}{\left\vert n-\left(
y-y_{1}\right) /a\right\vert ^{D+1}}\right] ,  \label{F2M}
\end{eqnarray}%
being the corresponding VEVs\ between two plates in the Minkowski bulk.
Here, $\zeta (x)$ is the Riemann zeta function. By taking into account that $%
\zeta (4)=\pi ^{4}/90$, for $D=3$ from (\ref{F2M}) we get%
\begin{equation}
\langle F^{2}\rangle _{\mathrm{M}}=\frac{1}{4\pi a^{4}}\left[ \pm
\sum_{n=-\infty }^{\infty }\frac{3\delta _{F}}{\left\vert
n-(y-y_{1})/a\right\vert ^{4}}-\frac{\pi ^{4}}{45}\right] .  \label{F2MD3}
\end{equation}%
Now, comparing with (\ref{F2D3b}) we see that for $D=3$ the last term in (%
\ref{F2D3b}) for the VEVs of the electric and magnetic fields squared is
conformally related to the corresponding VEV in the region between two
plates in the Minkowskian bulk. Near the plate $y=y_{1}$ the leading
contribution in (\ref{F2M}) comes from the term with $n=0$. As it has been
already mentioned, the corresponding leading term in the asymptotic
expansion over the distance from the plate coincides with that for the
Minkowski bulk (see (\ref{Enear}) for the electric field).

The last terms in (\ref{E21}) and (\ref{B21}) are induced by adding the
second plate to the geometry with a single plate at $z=z_{j}$. Let us
consider the asymptotic behavior of the second plate induced parts. In the
limit $z_{2}\rightarrow \infty $ for fixed $z$ and $z_{1}$ the dominant
contribution to the integrals in (\ref{E21}) and (\ref{B21}) with $j=1$
comes from the region near the lower limit of the integration. By using the
corresponding asymptotic expressions for the modified Bessel functions, for $%
\nu >0$ to the leading order we get%
\begin{eqnarray}
\langle F^{2}\rangle -\langle F^{2}\rangle _{1} &\approx &\frac{\delta _{\nu
}(D-1)(z/z_{2})^{D+2\nu }(z_{1}/z)^{2\nu (1-\delta _{\nu })}}{2^{D+2\nu
-3}\pi ^{D/2-1}\Gamma (D/2+1)\Gamma ^{2}(\nu )\alpha ^{D+1}}  \notag \\
&&\times \left[ 3-\delta _{\nu }-(1-\delta _{(F)})D\right] \int_{0}^{\infty
}dx\,x^{D+2\nu -1}\frac{K_{\nu }(x)}{I_{\nu }(x)}.  \label{E2B2as}
\end{eqnarray}%
In the case $D=3$ the asymptotic is directly obtained from (\ref{F2D3}):%
\begin{equation}
\langle F^{2}\rangle -\langle F^{2}\rangle _{1}\approx -\frac{\pi
^{3}(z/z_{2})^{4}}{180\alpha ^{4}}\left( 1-3\delta _{\nu }\delta
_{(F)}\right) .  \label{E2B2asD3}
\end{equation}%
For $D=4$ and for the boundary condition (\ref{BC2}) the corresponding
asymptotic has the form%
\begin{equation}
\langle F^{2}\rangle -\langle F^{2}\rangle _{1}\approx -\frac{3\delta
_{(F)}\left( z/z_{2}\right) ^{4}}{4\pi \alpha ^{5}\ln ^{2}(z_{1}/z_{2})}%
\int_{0}^{\infty }dx\,x^{3}\frac{K_{0}(x)}{I_{0}(x)}.  \label{E2B2asD4}
\end{equation}%
In terms of the physical distance of the observation point from the plate $%
z=z_{2}$, given by $y_{2}-y$, the second plate induced contributions decay
as $\exp [-(D+2\nu )(y_{2}-y)/\alpha ]$ for $D\geq 5$, as $%
e^{-4(y_{2}-y)/\alpha }/(y_{2}-y)^{2}$ for $D=4$ and like $%
e^{-4(y_{2}-y)/\alpha }$ for $D=3$. In all these cases one has an
exponential decay as a function of the distance. In the Minkowski bulk the
decay of the second plate induced contributions is power-law, like $%
1/(y_{2}-y)^{D+1}$.

In the limit $z_{1}\rightarrow 0$, for fixed values $z$ and $z_{2}$, the
corresponding asymptotics are obtained from (\ref{E21}) and (\ref{B21}) with
$j=2$ by taking into account that for $\nu >0$ one has%
\begin{equation}
\Omega _{\nu }^{(2)}(xz_{1},xz_{2})\approx \frac{2\nu (xz_{1}/2)^{2\nu }}{%
\Gamma ^{2}(\nu +1)I_{\nu }^{2}(xz_{2})}.  \label{Om2as}
\end{equation}%
As a consequence, the second boundary induced contributions $\langle
E^{2}\rangle -\langle E^{2}\rangle _{2}$ and $\langle B^{2}\rangle -\langle
B^{2}\rangle _{2}$ tend to zero like $(z_{1}/z_{2})^{2\nu }$. For $D=3$ the
corresponding asymptotic expressions are obtained from (\ref{F2D3}) and $%
\langle F^{2}\rangle -\langle F^{2}\rangle _{2}$ decays as $z_{1}/z_{2}$.
For $D=4$ and for the boundary condition (\ref{BC2}) one has $\nu =0$ and
the difference $\langle F^{2}\rangle -\langle F^{2}\rangle _{2}$ behaves as $%
1/\ln (z_{1}/z)$.

\begin{figure}[tbph]
\begin{center}
\begin{tabular}{cc}
\epsfig{figure=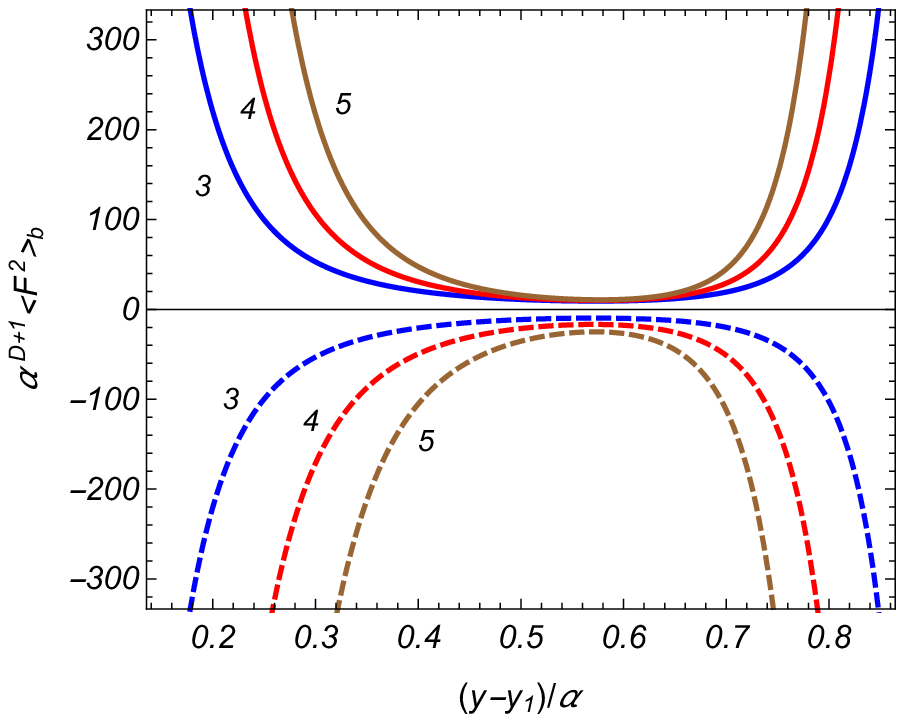,width=7.cm,height=5.5cm} & \quad %
\epsfig{figure=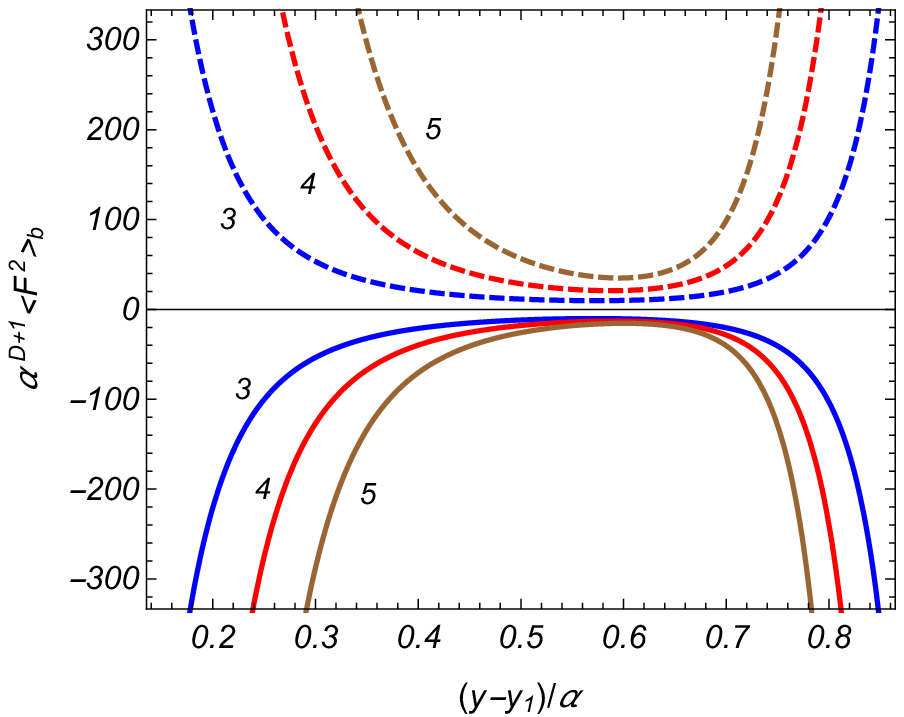,width=7.cm,height=5.5cm}%
\end{tabular}%
\end{center}
\caption{The boundary induced contributions in the VEVs of the electric
(full curves) and magnetic (dashed curves) fields squared in the region
between the plates as functions of the distance from the plate at $z=z_{1}$.
The left and right panels correspond to the boundary conditions (\protect\ref%
{BC}) and (\protect\ref{BC2}). The numbers near the curves are the values of
the spatial dimension $D$.}
\label{fig1}
\end{figure}

In figure \ref{fig1} we have plotted the boundary induced parts $\langle
F^{2}\rangle _{b}=\langle F^{2}\rangle -\langle F^{2}\rangle _{0}$ in the
VEVs of the electric ($F=E$) and magnetic ($F=B$) fields squared in the
region between the plates as functions of the distance from the plate at $%
z=z_{1}$ (in units of the curvature scale $\alpha $). The graphs are plotted
for $a=\alpha $ and the full and dashed curves correspond to the electric
and magnetic fields respectively. The numbers near the curves are the
corresponding values of the spatial dimension. The left and right panels are
plotted for the boundary conditions (\ref{BC}) and (\ref{BC2}). For the
example presented in figure \ref{fig1} the quantity $|\langle F^{2}\rangle
_{b}|$ increases with increasing $D$. This is not a general feature. For
example, in the case $a=2$ in the region near the point $y-y_{1}=a/2$ the
boundary induced VEV $|\langle F^{2}\rangle _{b}|$ decreases with increasing
$D$.

Having the correlators for the field strength tensor we can also evaluate
the Casimir-Polder forces acting on a polarizable particle for general case
of anisotropic polarizability tensor and dispersion. In the static limit and
for isoptropic polarizability $\alpha _{\mathrm{P}}$, the Casimir-Polder
potential is expressed in terms of the electric field squared as $U_{\mathrm{%
CP}}=-\alpha _{\mathrm{P}}\langle E^{2}\rangle /2$. From the graphs in
figure \ref{fig1} it follows that in the region between the plates the
Casimir-Polder forces near the plate at $z=z_{j}$ are attractive (repulsive)
with respect to that plate for the boundary condition (\ref{BC}) ((\ref{BC2}%
)). At some intermediate point one has $\partial _{z}\langle E^{2}\rangle =0$
and the force vanishes. This equilibrium position of a polarizable particle
is unstable for the condition (\ref{BC}) and stable for the condition (\ref%
{BC2}).

\section{VEV of the energy-momentum tensor}

\label{sec:emt}

Another important local characteristic of the vacuum state is the VEV of the
energy-momentum tensor:
\begin{equation}
\langle T_{\rho }^{\mu }\rangle =-\frac{1}{4\pi }\left[ \langle F_{\rho
\kappa }F^{\mu \kappa }\rangle -\frac{1}{4}\delta _{\rho }^{\mu }\langle
F_{\kappa \sigma }F^{\kappa \sigma }\rangle \right] .  \label{Tvev}
\end{equation}%
By using the expressions (\ref{FFc1}) and (\ref{FFc2}) for the two-point
functions in the coincidence limit, we can see that the off-diagonal
components vanish and for the diagonal components in the region $%
z_{1}<z<z_{2}$ we get
\begin{eqnarray}
\langle T_{\rho }^{\mu }\rangle &=&\langle T_{\rho }^{\mu }\rangle _{j}-%
\frac{\delta _{\nu }\delta _{\rho }^{\mu }(D-1)\alpha ^{-1-D}z^{D+2}}{%
2\left( 4\pi \right) ^{D/2}\Gamma (D/2+1)}\int_{0}^{\infty
}dx\,x^{D+1}\Omega _{\nu }^{(j)}(xz_{1},xz_{2})  \notag \\
&&\times \left[ \left( D-4\right) G_{\nu ,D/2-1}^{2}\left( xz_{j},xz\right)
+\left( D-2\right) G_{\nu ,D/2-2}^{2}\left( xz_{j},xz\right) \right] .
\label{Tmu}
\end{eqnarray}%
for $\mu ,\rho =0,\ldots ,D-1$ and%
\begin{eqnarray}
\langle T_{D}^{D}\rangle &=&\langle T_{D}^{D}\rangle _{j}-\frac{\delta _{\nu
}(D-1)\alpha ^{-1-D}z^{D+2}}{\left( 4\pi \right) ^{D/2}\Gamma (D/2)}%
\int_{0}^{\infty }dx\,x^{D+1}\Omega _{\nu }^{(j)}(xz_{1},xz_{2})  \notag \\
&&\times \left[ G_{\nu ,D/2-1}^{2}\left( xz_{j},xz\right) -G_{\nu
,D/2-2}^{2}\left( xz_{j},xz\right) \right] .  \label{TD}
\end{eqnarray}%
The expressions (\ref{Tmu}) and (\ref{TD}) with $j=1$ and $j=2$ provide two
equivalent representations of the VEVs. In these formulas, $\langle T_{\rho
}^{\mu }\rangle _{j}$ is the VEV\ in the geometry of a single plate located
at $z=z_{j}$. For $z>z_{j}$ one has%
\begin{eqnarray}
\langle T_{\rho }^{\mu }\rangle _{j} &=&\langle T_{\rho }^{\mu }\rangle _{0}-%
\frac{\delta _{\nu }\delta _{\rho }^{\mu }\left( D-1\right) \alpha
^{-1-D}z^{D+2}}{2\left( 4\pi \right) ^{D/2}\Gamma (D/2+1)}\int_{0}^{\infty
}dx\,x^{D+1}\frac{I_{\nu }(xz_{j})}{K_{\nu }(xz_{j})}  \notag \\
&&\times \left[ \left( D-4\right) K_{D/2-1}^{2}(xz)+\left( D-2\right)
K_{D/2-2}^{2}(xz)\right] ,  \notag \\
\langle T_{D}^{D}\rangle _{j} &=&\langle T_{D}^{D}\rangle _{0}-\frac{\delta
_{\nu }\left( D-1\right) \alpha ^{-1-D}z^{D+2}}{\left( 4\pi \right)
^{D/2}\Gamma (D/2)}\int_{0}^{\infty }dx\,x^{D+1}\,\frac{I_{\nu }(xz_{j})}{%
K_{\nu }(xz_{j})}  \notag \\
&&\times \left[ K_{D/2-1}^{2}(xz)-K_{D/2-2}^{2}(xz)\right] ,  \label{Tj1}
\end{eqnarray}%
with $\nu $ given by (\ref{nu}). In the region $z<z_{j}$ the VEVs $\langle T_{\nu }^{\mu }\rangle _{j}$ are
obtained from (\ref{Tj1}) by the replacements $I\rightleftarrows K$ of the
modified Bessel functions in the boundary-induced contributions. In the
region $z<z_{1}$ the VEV\ of the energy-momentum tensor is given by (\ref%
{Tj1}) with $j=1$ and with the replacements $I\rightleftarrows K$. The VEV
in the region $z>z_{2}$ is given by (\ref{Tj1}) with $j=2$. The
boundary-induced VEVs in the geometry of a single plate and for the boundary
condition (\ref{BC}) have been previously investigated in \cite{Kota16}-\cite%
{Kota17b}. For the same boundary condition, a part of the results concerning
the VEV of the energy-momentum tensor in the region between the plates has
been presented at the 10th Alexander Friedmann International Conference \cite%
{Saha20}. As seen from (\ref{Tmu}), the components $\langle T_{\rho }^{\mu
}\rangle $ with $\mu ,\rho =0,1,\ldots ,D-1$, coincide. This is a
consequence of the Poincar\'{e} invariance of the problem in the subspace $%
(x^{0},x^{1},\ldots ,x^{D-1})$. In the expressions (\ref{Tmu})-(\ref{Tj1})
and for points away from the plates, the renormalization is required for the
boundary-free contribution $\langle T_{\rho }^{\mu }\rangle _{0}$ only. From
the maximal symmetry of the bulk geometry we expect that the corresponding
renormalized VEV has the form $\langle T_{\rho }^{\mu }\rangle _{0}=\mathrm{%
const}\cdot $ $\delta _{\rho }^{\mu }$ and it does not depend on spacetime
point.

We will denote by $\langle T_{\rho }^{\mu }\rangle _{\mathrm{b}}=\langle
T_{\rho }^{\mu }\rangle -\langle T_{\rho }^{\mu }\rangle _{0}$ the boundary
induced contribution to the VEV of the energy-momentum tensor. One can check
that it obeys the covariant continuity equation $\nabla _{\mu }\langle
T_{\rho }^{\mu }\rangle _{\mathrm{b}}=0$. In the problem under consideration
the latter is reduced to the relation
\begin{equation}
z^{D+1}\partial _{z}\left( z^{-D}\langle T_{D}^{D}\rangle _{\mathrm{b}%
}\right) +D\langle T_{0}^{0}\rangle _{\mathrm{b}}=0  \label{Conteq}
\end{equation}%
between the energy density and the normal stress. For $D>3$ the boundary
induced contribution in the vacuum energy density is negative for the
boundary condition (\ref{BC}) and positive for the boundary condition (\ref%
{BC2}) for all values of $z$, including the regions $z<z_{1}$, $z>z_{2}$.
Near the plate $z=z_{j}$ and for the boundary condition (\ref{BC})/(\ref{BC2}%
), the boundary induced contribution to the normal stress $\langle
T_{D}^{D}\rangle $ is negative/positive in the region $z>z_{j}$ and
positive/negative in the region $z<z_{j}$. The trace of the boundary induced
contribution in the VEV of the energy-momentum tensor is related to the
corresponding photon condensate by the formula%
\begin{equation}
\langle T_{\mu }^{\mu }\rangle _{\mathrm{b}}=-\frac{D-3}{16\pi }\left\langle
F_{\mu \sigma }F^{\mu \sigma }\right\rangle _{\mathrm{b}}.  \label{Trace}
\end{equation}%
For $D=3$ that contribution is traceless. The latter property is a
consequence of the conformal invariance of the electromagnetic field in $D=3$%
. Note that because of the conformal anomaly the boundary free part has
nonzero trace, $\langle T_{\mu }^{\mu }\rangle _{0}\neq 0$.

For $D>3$ the VEVs diverge on the plates. The divergence on the plate $%
z=z_{j}$ comes from the single boundary part $\langle T_{\rho }^{\mu
}\rangle _{j}$. The leading terms in the asymptotic expansion over the
distance from the plate are expressed as
\begin{eqnarray}
\langle T_{0}^{0}\rangle &\approx &-\frac{D-3}{12\pi }\langle E^{2}\rangle ,
\notag \\
\langle T_{D}^{D}\rangle &\approx &(z/z_{j}-1)\langle T_{0}^{0}\rangle
\approx \frac{y-y_{j}}{\alpha }\langle T_{0}^{0}\rangle ,  \label{T00near}
\end{eqnarray}%
with $\langle E^{2}\rangle $ given by (\ref{Enear}). This type of divergence
is well-known in quantum field theory with boundaries. In (\ref{Tmu}) and (%
\ref{TD}) the second plate induced contributions (the second terms in the
right-hand sides) are finite at $z=z_{j}$.

In the special case $D=3$, by using the expressions for the functions $%
I_{\pm 1/2}(u)$, $K_{\pm 1/2}(u)$, and (\ref{G12}) we can see that in the
region between the plates
\begin{eqnarray}
\langle T_{\rho }^{\mu }\rangle &=&\langle T_{\rho }^{\mu }\rangle _{0}-%
\frac{\delta _{\rho }^{\mu }\pi ^{2}\left( z/\alpha \right) ^{4}}{720\left(
z_{2}-z_{1}\right) ^{4}},  \notag \\
\langle T_{3}^{3}\rangle &=&\langle T_{3}^{3}\rangle _{0}+\frac{\pi
^{2}\left( z/\alpha \right) ^{4}}{240\left( z_{2}-z_{1}\right) ^{4}},
\label{TmuD3}
\end{eqnarray}%
where $\mu \neq 3$. Note that the boundary-induced contributions (the second
terms in the right-hand sides) in these expressions are the same for the
boundary conditions (\ref{BC}) and (\ref{BC2}). Those contributions are
conformally related to the corresponding VEVs in the region between two
plates in the Minkowski bulk with the conformal factor $\left( z/\alpha
\right) ^{4}$. In the region $z>z_{2}$ one has $\langle T_{\rho }^{\mu
}\rangle =\langle T_{\rho }^{\mu }\rangle _{0}$. In the region $z<z_{1}$ the
VEVs for the boundary condition (\ref{BC}) are given by the expressions ($%
\mu ,\rho =0,1,2$)%
\begin{eqnarray}
\langle T_{\rho }^{\mu }\rangle &=&\langle T_{\rho }^{\mu }\rangle _{0}-%
\frac{\delta _{\rho }^{\mu }\pi ^{2}}{720}\left( \frac{z}{\alpha z_{1}}%
\right) ^{4},  \notag \\
\langle T_{3}^{3}\rangle &=&\langle T_{3}^{3}\rangle _{0}+\frac{\pi ^{2}}{240%
}\left( \frac{z}{\alpha z_{1}}\right) ^{4}.  \label{TmuD3L}
\end{eqnarray}%
From (\ref{TmuD3L}) it follows that the $z$-projection of Casimir force
acting per unit surface of the plate at $z=z_{1}$ (from the side $z=z_{1}-0$%
) is given by $-\pi ^{2}\alpha ^{-4}/240$. The latter does not depend on the
location of the plate and is directed towards the AdS boundary. For the
boundary condition (\ref{BC2}) the expressions for VEVs in the region $%
z<z_{1}$ read
\begin{eqnarray}
\langle T_{\rho }^{\mu }\rangle &=&\langle T_{\rho }^{\mu }\rangle
_{0}+\delta _{\rho }^{\mu }\frac{7\pi ^{2}}{5760}\left( \frac{z}{\alpha z_{1}%
}\right) ^{4},  \notag \\
\langle T_{3}^{3}\rangle &=&\langle T_{3}^{3}\rangle _{0}-\frac{7\pi ^{2}}{%
1920}\left( \frac{z}{\alpha z_{1}}\right) ^{4},  \label{TDDcL}
\end{eqnarray}%
with $\mu =0,1,2$. Note that (\ref{TDDcL}) differs from the corresponding
result for the boundary condition (\ref{BC}) (given by (\ref{TmuD3L})). The
corresponding energy density is positive and the $z$-projection of Casimir
force acting per unit surface of the side $z=z_{1}-0$ is given by $7\pi
^{2}\alpha ^{-4}/1920$. The latter is repulsive with respect to the AdS
boundary.

In the Minkowskian limit $\alpha \rightarrow \infty $ for fixed $y$, $y_{j}$%
, by using the relations (\ref{OmMink}), to the leading order we get $%
\langle T_{\rho }^{\mu }\rangle \rightarrow \langle T_{\rho }^{\mu }\rangle
_{\mathrm{M}}$, with the Minkowskian VEVs%
\begin{eqnarray}
\langle T_{\rho }^{\mu }\rangle _{\mathrm{M}} &=&-\delta _{\rho }^{\mu }%
\frac{(D-1)\Gamma ((D+1)/2)}{\left( 4\pi \right) ^{(D+1)/2}a^{D+1}}\left[
\zeta (D+1)\pm \sum_{n=-\infty }^{\infty }\frac{(D-3)/2}{\left\vert n-\left(
y-y_{1}\right) /a\right\vert ^{D+1}}\right] ,  \notag \\
\langle T_{D}^{D}\rangle _{\mathrm{M}} &=&\frac{D(D-1)\Gamma ((D+1)/2)}{%
\left( 4\pi \right) ^{(D+1)/2}a^{D+1}}\zeta (D+1),  \label{TDDM}
\end{eqnarray}%
where $\mu =0,1,\ldots ,D-1$, and the upper and lower signs correspond to
the boundary conditions (\ref{BC}) and (\ref{BC2}). For $D=3$ these results
are conformally related to the boundary-induced VEVs in the AdS bulk (see (%
\ref{TmuD3})). Note that the normal stress is uniform and is the same for
the conditions (\ref{BC}) and (\ref{BC2}).

The last terms in (\ref{Tmu}) and (\ref{TD}) are the contribution induced by
the second plate when we add it to the geometry with a single plate located
at $z=z_{j}$. Let us consider the corresponding asymptotics for limiting
values of the second plate location. In the limit $z_{2}\rightarrow \infty $%
, when $z$ and $z_{1}$ are fixed, for $D>4$ from (\ref{Tmu}) and (\ref{TD})
with $j=1$ we find%
\begin{eqnarray}
\langle T_{D}^{D}\rangle -\langle T_{D}^{D}\rangle _{1} &\approx &-\delta
_{\nu }D\frac{\langle T_{0}^{0}\rangle -\langle T_{0}^{0}\rangle _{1}}{%
D-3+\delta _{\nu }}\approx \frac{(D-1)\alpha ^{-1-D}\left( z/z_{2}\right)
^{D+2\nu }}{2^{D+2\nu -2}\pi ^{D/2}\Gamma (D/2)\Gamma ^{2}(\nu )}  \notag \\
&&\times (z_{1}/z)^{2\nu \left( 1-\delta _{\nu }\right) }\int_{0}^{\infty
}dx\,x^{D+2\nu -1}\frac{K_{\nu }(x)}{I_{\nu }(x)},  \label{TDDas}
\end{eqnarray}%
and the second plate contributions decay as $(z/z_{2})^{D+2\nu }$. Note that
for the boundary condition (\ref{BC}) the leading term does not depend on $%
z_{1}$. For $D=4$, the leading terms in the limit $z_{2}\rightarrow \infty $
are given by
\begin{equation}
\langle T_{D}^{D}\rangle -\langle T_{D}^{D}\rangle _{1}\approx -2\left(
\langle T_{0}^{0}\rangle -\langle T_{0}^{0}\rangle _{1}\right) \approx \frac{%
3\alpha ^{-5}\left( z_{1}/z_{2}\right) ^{4}}{16\pi ^{2}\ln ^{2}(z_{1}/z_{2})}%
\left( z/z_{1}\right) ^{6}\int_{0}^{\infty }dx\,x^{3}\frac{K_{0}(x)}{I_{0}(x)%
}.  \label{TDDasD4}
\end{equation}%
The asymptotic for $D=3$ is directly obtained from (\ref{TmuD3}). The
leading terms in the limit $z_{1}\rightarrow 0$, for fixed $z$ and $z_{2}$,
are obtained from (\ref{Tmu}) and (\ref{TD}) with $j=2$ by the replacement (%
\ref{Om2as}) and the second plate induced contributions vanish like $%
(z_{1}/z_{2})^{D-2}$.

\begin{figure}[tbph]
\begin{center}
\begin{tabular}{cc}
\epsfig{figure=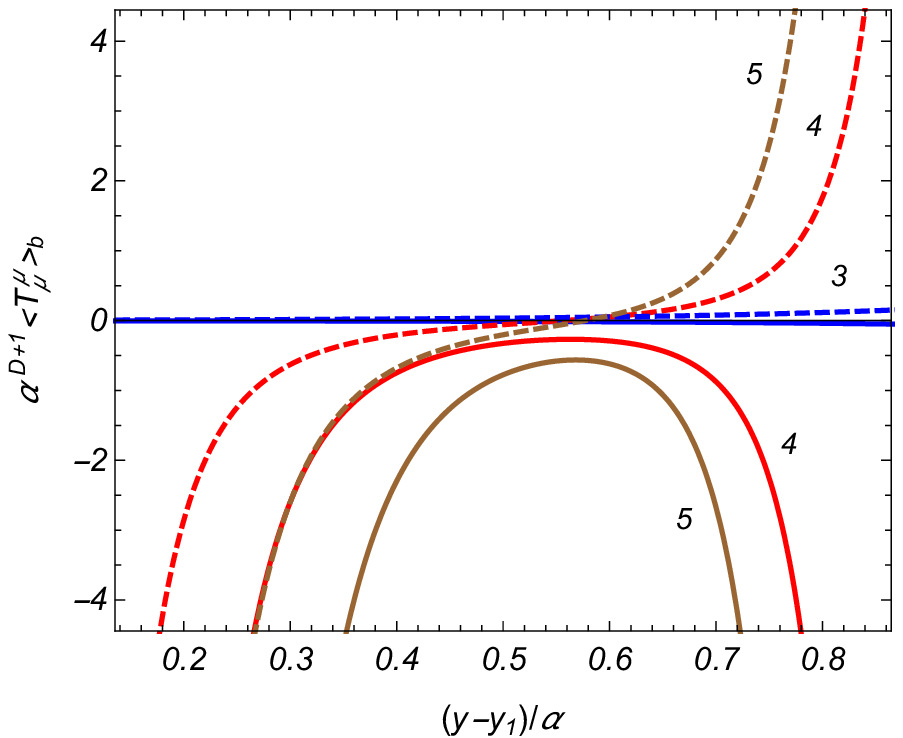,width=7.cm,height=5.5cm} & \quad %
\epsfig{figure=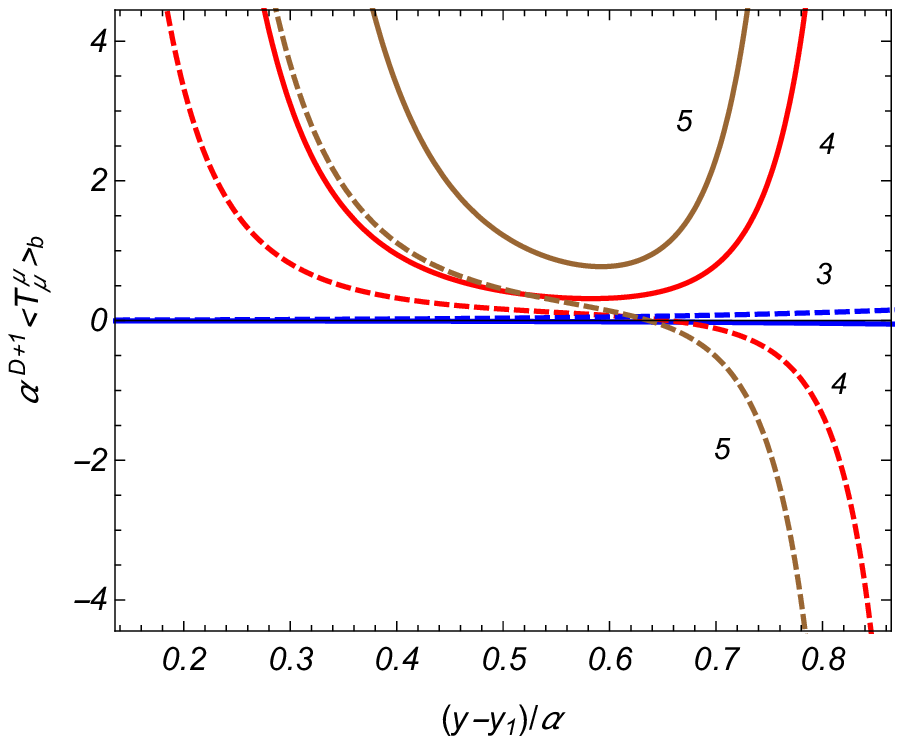,width=7.cm,height=5.5cm}%
\end{tabular}%
\end{center}
\caption{The boundary induced parts in the VEVs of the energy density (full
curves) and the normal stress (dashed curves) versus the distance from the
plate at $z=z_{1}$. The left and right panels present the graphs for the
boundary conditions (\protect\ref{BC}) and (\protect\ref{BC2}),
respectively. The numbers near the curves correspond to the values of $D$.}
\label{fig2}
\end{figure}

Figure \ref{fig2} presents the boundary induced contributions in the VEVs of
the energy density ($\mu =0$, full curves) and the normal stress ($\mu =D$,
dashed curves) in the region between the plates with the separation $%
a/\alpha =1$. The left and right panels correspond to the boundary
conditions (\ref{BC}) and (\ref{BC2}), respectively. For $D=3$ the VEVs are
finite on the boundaries and in the region between the plates they coincide
for the conditions (\ref{BC}) and (\ref{BC2}).

\section{The Casimir forces}

\label{sec:Force}

The Casimir force acting on the plate is determined by the normal stress $%
\langle T_{D}^{D}\rangle $ evaluated at the location of the plate. The
boundary-free contributions to the normal stress are the same on the left-
and right-hand sides of the plate and will not contribute to the net force.
For the plate at $z=z_{j}$, the remaining contribution to the Casimir force
per unit surface is decomposed into two parts:%
\begin{equation}
p_{j}=p_{j}^{\mathrm{(s)}}+p_{j}^{\mathrm{(int)}},  \label{pj}
\end{equation}%
where $p_{j}^{\mathrm{(s)}}$ is the vacuum pressure on the plate at $z=z_{j}$
when the second plate is absent (self-action force) and the part $p_{j}^{%
\mathrm{(int)}}$ is induced by the second plate (interaction force). The
expression for $p_{j}^{\mathrm{(s)}}$ is obtained by combining the vacuum
pressures on the left- and right-hand sides of the plate. The $z$-projection
of the self-action force acting per unit surface of the plate at $z=z_{j}$
is given by $f_{j}^{\mathrm{(s)}}=\langle T_{D}^{D}\rangle
_{j}|_{z=z_{j}-0}^{z=z_{j}+0}$. For $D>3$ the self-action contributions are
divergent and they require an additional renormalization. For $D=3$ and for
the boundary condition (\ref{BC}) one has $\langle T_{D}^{D}\rangle
_{j}=\langle T_{D}^{D}\rangle _{0}$ for $z=z_{j}+0$ and $\langle
T_{D}^{D}\rangle _{j}=\langle T_{D}^{D}\rangle _{0}+\pi ^{2}\alpha ^{-4}/240$
for $z=z_{j}-0$. Hence, in this case one gets%
\begin{equation}
f_{j}^{\mathrm{(s)}}=-\frac{\pi ^{2}}{240\alpha ^{4}},  \label{fjsD3}
\end{equation}%
and the self-action force is directed towards the AdS boundary. In a similar
way, for the boundary condition (\ref{BC2}) and for $D=3$ we find%
\begin{equation}
f_{j}^{\mathrm{(s)}}=\frac{7\pi ^{2}}{1920\alpha ^{4}}.  \label{fjsD3b}
\end{equation}%
For this case the force is directed from the AdS boundary. Note that in both
cases of the boundary conditions the self-action force per unit surface do
not depend on the location of the plate.

In contrast to the self-action part, the interaction term $p_{j}^{\mathrm{%
(int)}}$ is finite and does not require a further renormalization. By using
the expression (\ref{TD}) for the normal stress, one finds%
\begin{eqnarray}
p_{1}^{\mathrm{(int)}} &=&-\frac{(D-1)\alpha ^{-1-D}}{2^{D}\pi ^{D/2}\Gamma
(D/2)}\int_{0}^{\infty }dx\,\frac{x^{D-1}K_{\nu }(x\eta )}{K_{\nu }(x)G_{\nu
,\nu }(x,x\eta )},  \notag \\
p_{2}^{\mathrm{(int)}} &=&-\frac{(D-1)\alpha ^{-1-D}}{2^{D}\pi ^{D/2}\Gamma
(D/2)}\int_{0}^{\infty }dx\,\frac{x^{D-1}I_{\nu }(x/\eta )}{I_{\nu
}(x)G_{\nu ,\nu }(x/\eta ,x)},  \label{p12}
\end{eqnarray}%
where $\eta $ is defined as (\ref{eta}). Here, $p_{1}^{\mathrm{(int)}}$ acts
on the side $z=z_{1}+0$ for the plate at $z=z_{1}$ and $p_{2}^{\mathrm{(int)}%
}$ acts on the side $z=z_{2}-0$ for the plate at $z=z_{2}$. The $z$%
-projection of the corresponding force acting per unit surface of the plate
at $z=z_{j}$ is given by $f_{j}^{\mathrm{(int)}}=(-1)^{j}p_{j}^{\mathrm{(int)%
}}$. By taking into account that $G_{\nu ,\nu }(u,v)>0$ for $v>u$, we see
that $p_{j}^{\mathrm{(int)}}<0$ and the interaction forces between the
plates are always attractive. As is seen, the interaction terms depend on
the locations of the plates through the ratio $z_{2}/z_{1}$. This is a
consequence of the maximal symmetry of the AdS spacetime. Note that the
integrand in (\ref{p12}) for $p_{j}^{\mathrm{(int)}}$ is also expressed as $%
x^{D-1}\Omega _{\nu }^{(j)}(x,x\eta )$. By taking into account that
\begin{equation}
\Omega _{\nu }^{(j)}(uz_{1},uz_{2})=(-1)^{j}z_{j}\partial _{z_{j}}\ln
\left\vert 1-\frac{I_{\nu }(uz_{1})K_{\nu }(uz_{2})}{I_{\nu }(uz_{2})K_{\nu
}(uz_{1})}\right\vert ,  \label{Omder}
\end{equation}%
the $z$-projection of the interaction force is presented in an alternative
form%
\begin{equation}
f_{j}^{\mathrm{(int)}}=-\frac{(D-1)\alpha ^{-1-D}z_{j}^{D+1}}{2^{D}\pi
^{D/2}\Gamma (D/2)}\partial _{z_{j}}\int_{0}^{\infty }du\,u^{D-1}\ln
\left\vert 1-\frac{I_{\nu }(uz_{1})K_{\nu }(uz_{2})}{I_{\nu }(uz_{2})K_{\nu
}(uz_{1})}\right\vert .  \label{fja}
\end{equation}

\begin{figure}[tbph]
\begin{center}
\epsfig{figure=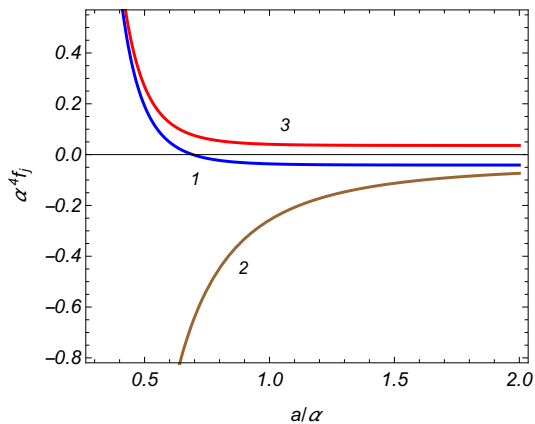,width=7.cm,height=5.5cm}
\end{center}
\caption{The $z$-projections of the Casimir forces for $D=3$ as functions of
the separation between the plates. The curves 1 and 3 correspond to the
force $f_{1}$ for the boundary conditions (\protect\ref{BC}) and (\protect
\ref{BC2}), respectively. The curve 2 correspond to the force $f_{2}$ (it
coincides for the conditions (\protect\ref{BC}) and (\protect\ref{BC2}).}
\label{fig3}
\end{figure}

In the Minkowskian limit, $\alpha \rightarrow \infty $, by using the
asymptotic expressions (\ref{OmMink}), we can see that $p_{j}^{\mathrm{(int)}%
}\rightarrow p_{\mathrm{M}}$, where the Casimir pressure in the Minkowski
bulk is given by
\begin{equation}
p_{\mathrm{M}}=-D(D-1)\frac{\Gamma \left( \left( D+1\right) /2\right) \zeta
(D+1)}{\left( 4\pi \right) ^{(D+1)/2}a^{D+1}}.  \label{pMj}
\end{equation}%
Note that the self-action stresses $\langle T_{D}^{D}\rangle
_{j}|_{z=z_{j}-0}$ and $\langle T_{D}^{D}\rangle _{j}|_{z=z_{j}+0}$ on a
single plate in the Minkowski bulk are equal and the corresponding net force
vanishes. Another special case corresponds to $D=3$ with general $\alpha $.
The interaction parts are obtained from (\ref{p12}). The $z$-projection of
the total force acting per unit surface of the plate at $z=z_{j}$ is
expressed as%
\begin{equation}
f_{j}=f_{j}^{\mathrm{(s)}}+(-1)^{j}p_{j}^{\mathrm{(int)}}.  \label{fj}
\end{equation}%
By taking into account the expressions (\ref{fjsD3}) and (\ref{fjsD3b}) for $%
f_{j}^{\mathrm{(s)}}$, for the boundary condition (\ref{BC}) one finds
\begin{eqnarray}
f_{1} &=&-\frac{\pi ^{2}}{240\alpha ^{4}}\left[ 1-\left( e^{a/\alpha
}-1\right) ^{-4}\right] ,  \notag \\
f_{2} &=&-\frac{\pi ^{2}}{240\alpha ^{4}}\,\left( 1-e^{-a/\alpha }\right)
^{-4},  \label{f12BC}
\end{eqnarray}%
where $a/\alpha $ is the proper distance between the plates measured in
units of the AdS curvature scale $\alpha $. For the boundary condition (\ref%
{BC2}) $f_{2}$ coincides with (\ref{f12BC}) and for $f_{1}$ we get
\begin{equation}
f_{1}=\frac{\pi ^{2}}{240\alpha ^{4}}\left[ \frac{7}{8}+\left( e^{a/\alpha
}-1\right) ^{-4}\right] .  \label{f2BC2}
\end{equation}%
The difference of the forces $f_{1}$ for the boundary conditions (\ref{BC})
and (\ref{BC2}) is a consequence of different interactions of the plate at $%
z=z_{1}$ with the AdS boundary. In figure \ref{fig3} we have plotted the
forces $f_{j}$ from (\ref{f12BC}) and (\ref{f2BC2}) as functions of $%
a/\alpha $. The curve 1 corresponds to the force $f_{1}$ for the condition (%
\ref{BC}), the curve 2 presents the force $f_{2}$ (is the same for both the
boundary conditions), and the curve 3 corresponds to the force $f_{1}$ for
the condition (\ref{BC2}). As seen from the graphs, the Casimir force on the
plate at $z=z_{2}$ is always directed toward the AdS boundary for both
boundary conditions. For the boundary condition (\ref{BC2}) the force on the
plate at $z=z_{1}$ is directed toward the AdS horizon. In the case of the
boundary condition (\ref{BC}) the force acting on the plate at $z=z_{1}$ is
directed toward the AdS horizon for small separations between the plates and
towards the AdS boundary for large separations. At $a/\alpha \approx 0.693$
that force becomes zero.

Now let us consider the asymptotics of the interaction forces at small and
large separations between the plates. For proper distances much smaller than
the AdS curvature scale one has $a/\alpha \ll 1$ and $z_{2}/z_{1}-1\ll 1$.
In this limit the dominant contribution to the integrals in (\ref{p12}) come
from large values of $x$. By using the asymptotic expressions for the
modified Bessel functions for large arguments we can see that to the leading
order $p_{1}^{\mathrm{(int)}}\approx p_{2}^{\mathrm{(int)}}\approx p_{%
\mathrm{M}}$, where the vacuum pressures in the Minkowski bulk are given by (%
\ref{pMj}). This shows that for small separations the effect of gravity on
the Casimir forces is weak. This is related to fact that at such separations
the dominant contribution to the forces come from the vacuum fluctuations
with wavelengths smaller than the curvature radius. The influence of the
gravitational field on these fluctuations is weak. At large separations
between the plates one has $a/\alpha \gg 1$ and $z_{2}/z_{1}\gg 1$. For $D=3$
the corresponding asymptotics are obtained from (\ref{f12BC}) and (\ref%
{f2BC2}). For $D>3$ from (\ref{p12}) to the leading order we find%
\begin{eqnarray}
p_{1}^{\mathrm{(int)}} &\approx &-\frac{4(D-1)\alpha ^{-1-D}e^{-(D+2\nu
)a/\alpha }}{2^{D+2\nu }\pi ^{D/2}\Gamma (D/2)\Gamma ^{2}(\nu )}%
\int_{0}^{\infty }dx\,x^{D+2\nu -1}\frac{K_{\nu }(x)}{I_{\nu }(x)},  \notag
\\
p_{2}^{\mathrm{(int)}} &\approx &-\frac{2\nu (D-1)\alpha ^{-1-D}e^{-2\nu
a/\alpha }}{2^{D+2\nu }\pi ^{D/2}\Gamma (D/2)\Gamma ^{2}(\nu +1)}%
\int_{0}^{\infty }dx\,\frac{x^{D+2\nu -1}}{I_{\nu }^{2}(x)}.
\label{LargeDist}
\end{eqnarray}%
The case $D=4$ for the boundary condition (\ref{BC2}) should be considered
separately:%
\begin{eqnarray}
p_{1}^{\mathrm{(int)}} &\approx &-\frac{3e^{-4a/\alpha }}{16\pi ^{2}\alpha
^{3}a^{2}}\int_{0}^{\infty }dx\,x^{3}\frac{K_{0}(x)}{I_{0}(x)},  \notag \\
p_{2}^{\mathrm{(int)}} &\approx &-\frac{3\alpha ^{-4}}{16\pi ^{2}a}%
\int_{0}^{\infty }dx\,\frac{x^{3}}{I_{0}^{2}(x)}.  \label{p2LD4}
\end{eqnarray}%
The integrals in (\ref{p2LD4}) are equal to 1.0045 and 4.0181 for $p_{1}^{%
\mathrm{(int)}}$ and $p_{2}^{\mathrm{(int)}}$, respectively. From (\ref%
{LargeDist}) we see that for confining boundary conditions (\ref{BC2}) the
decay of the interaction forces is weaker. For both the boundary conditions (%
\ref{BC}) and (\ref{BC2}) at large separations the interaction parts are
exponentially suppressed as functions of the separation $a$. This behavior
is in clear contrast with that for the Minkowski bulk where the Casimir
forces decay as $1/a^{D+1}$ for all separations between the plates.

\begin{figure}[tbph]
\begin{center}
\begin{tabular}{cc}
\epsfig{figure=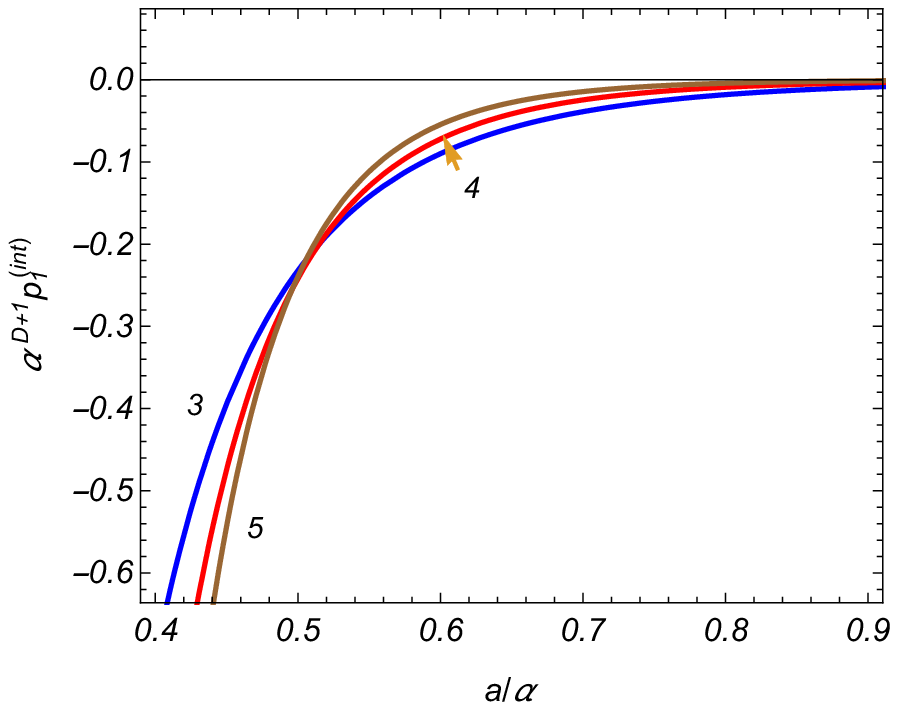,width=7.cm,height=5.5cm} & \quad %
\epsfig{figure=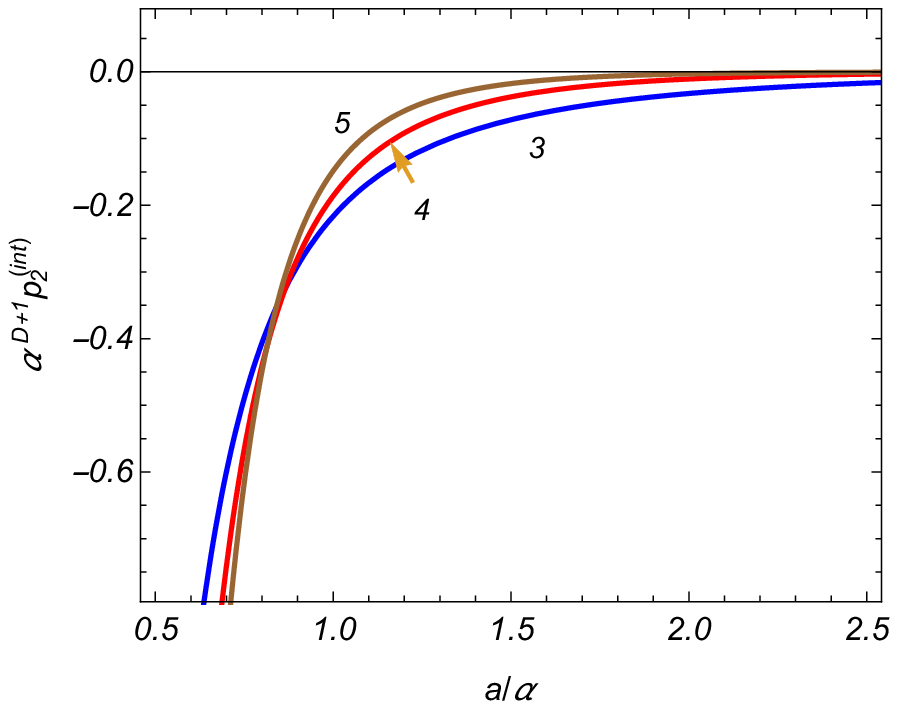,width=7.cm,height=5.5cm}%
\end{tabular}%
\end{center}
\caption{The interaction parts in the vacuum pressures on the plates for the
boundary condition (\protect\ref{BC}) as functions of the separation. The
left and right panels correspond to the forces acting on the plates $z=z_{1}$
and $z=z_{2}$, respectively, and the numbers near the curves are the values
of the spatial dimension $D$.}
\label{fig4}
\end{figure}

\begin{figure}[tbph]
\begin{center}
\begin{tabular}{cc}
\epsfig{figure=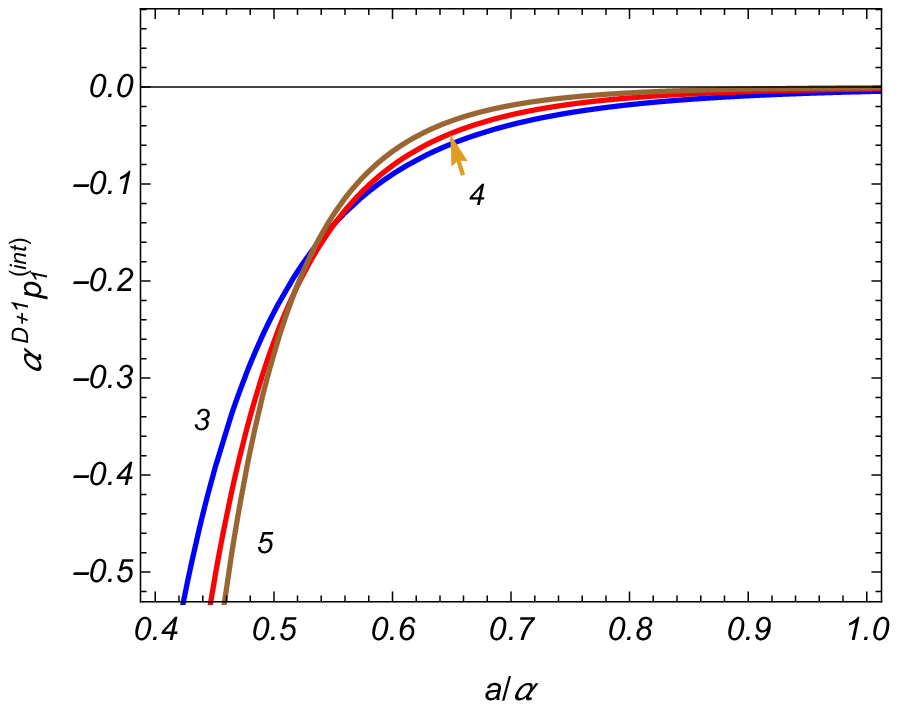,width=7.cm,height=5.5cm} & \quad %
\epsfig{figure=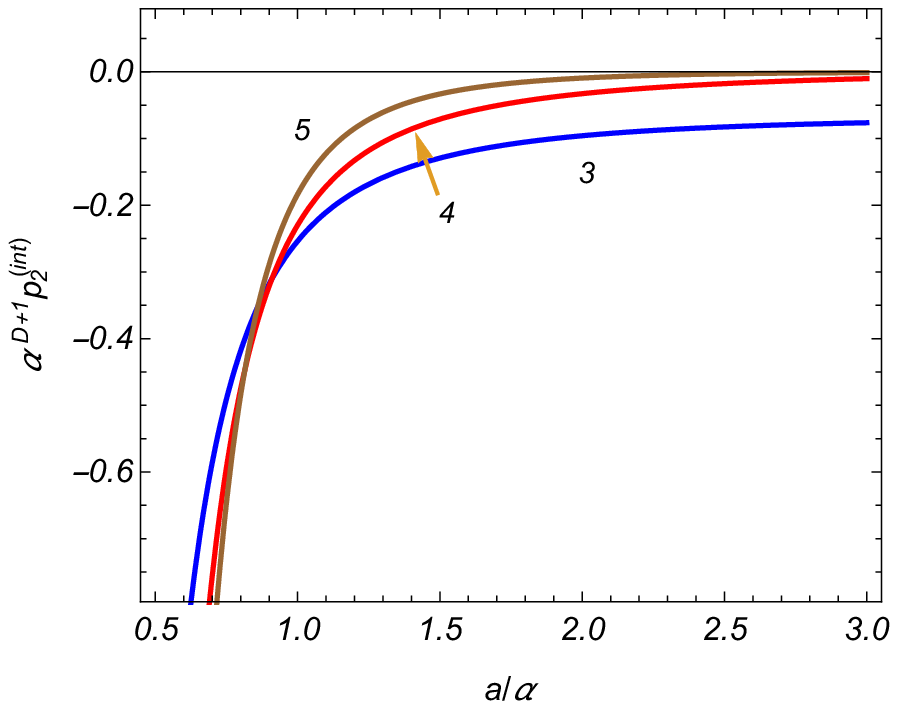,width=7.cm,height=5.5cm}%
\end{tabular}%
\end{center}
\caption{The same as in figure \protect\ref{fig4} for the boundary condition
(\protect\ref{BC2}).}
\label{fig5}
\end{figure}

In figures \ref{fig4} and \ref{fig5} we display the interaction pressures (%
\ref{p12}) versus the separation between the plates (in units of the
curvature radius) for different values of the spatial dimension $D$ (numbers
near the curves). The left and right panels correspond to the forces acting
on the plates $z=z_{1}$ and $z=z_{2}$, respectively. Figure \ref{fig4} is
plotted for the boundary condition (\ref{BC}) and figure \ref{fig5} presents
the results for the condition (\ref{BC2}).

\section{Vacuum energy}

\label{sec:VacEn}

In the previous sections we have considered the local characteristics of the
vacuum and the Casimir forces acting on the boundaries. Here we are
interested in the total vacuum energy and its relation to the Casimir forces.

\subsection{Zeta function and the vacuum energy in the region between the
plates}

In the region between the plates the corresponding eigenvalues of the
quantum number $\lambda $ are solutions of the equation (\ref{lameig}) where
$\nu $ is defined as (\ref{nu}) and the function $g_{\nu ,\rho }(x,y)$ is
given by (\ref{genu}). First let us consider the expression for the VEV of
the energy density based on the representation (\ref{AA1}) for the two-point
function. By using (\ref{Tvev}) we get%
\begin{eqnarray}
\langle T_{0}^{0}\rangle &=&\frac{\left( D-1\right) \alpha ^{-1-D}z^{D+2}}{%
16\left( 2\pi \right) ^{D-3}z_{1}^{2}}\int d\mathbf{k}\,\sum_{n=1}^{\infty }%
\frac{u}{\omega _{\nu ,n}}T_{\nu }(\eta ,u)\Big\{\omega _{\nu ,n}^{2}g_{\nu
,D/2-1}^{2}\left( u,uz/z_{1}\right)  \notag \\
&&+\left( \frac{k^{2}}{D-1}+\frac{u^{2}}{2z_{1}^{2}}\right) \left[ g_{\nu
,D/2-2}^{2}\left( u,uz/z_{1}\right) -g_{\nu ,D/2-1}^{2}\left(
u,uz/z_{1}\right) \right] \big\}_{u=\lambda _{\nu ,n}},  \label{T00Rep2}
\end{eqnarray}%
where $\omega _{\nu ,n}$ is given by (\ref{omeq}).

By taking into account the boundary conditions (\ref{lameig}) and making use
of the integral for the square of cylinder functions, it can be seen that%
\begin{equation*}
\int_{z_{1}}^{z_{2}}dz\,zg_{\nu
,D/2-1}^{2}(u,uz/z_{1})=\int_{z_{1}}^{z_{2}}dz\,zg_{\nu
,D/2-2}^{2}(u,uz/z_{1})=\frac{2z_{1}^{2}}{\pi ^{2}uT_{\nu }(\eta ,u)},
\end{equation*}%
where $u=\lambda _{\nu ,n}$. Combining this with (\ref{T00Rep2}) the
following relation is obtained
\begin{equation}
\mathcal{E}_{\mathrm{(II)}}=\int_{z_{1}}^{z_{2}}dz\,\left( \frac{\alpha }{z}%
\right) ^{D+1}\langle T_{0}^{0}\rangle .  \label{Enint}
\end{equation}%
Here%
\begin{equation}
\mathcal{E}_{\mathrm{(II)}}=\frac{1}{2}\sum_{\sigma }\int \frac{d\mathbf{k}}{%
(2\pi )^{D-1}}\,\sum_{n=1}^{\infty }\omega _{\nu ,n}=\frac{D-1}{2}\int \frac{%
d\mathbf{k}}{(2\pi )^{D-1}}\sum_{n=1}^{\infty }\sqrt{\lambda _{\nu
,n}^{2}/z_{1}+k^{2}},  \label{Evac}
\end{equation}%
is the total vacuum energy in the region $z_{1}\leq z\leq z_{2}$ (region II)
per unit coordinate volume along the parallel directions $(x^{1},\ldots
,x^{D-1})$. The factor $D-1$ in the second expression is the number of
independent polarizations of the electromagnetic field in $D$-dimensional
space. Hence, we have seen that the integral of the bulk energy density is
equal to the vacuum energy evaluated as the sum of the ground state energies
of elementary oscillators. Note that these two quantities, in general, can
be different. The difference may be related to the surface energy density
located on the boundaries. An example of this kind of problem is provided by
a scalar field with Robin boundary condition \cite{Rome02}. The
corresponding surface energy-momentum tensor for general bulk and boundary
geometries has been considered in \cite{Saha04}. The VEV\ of the surface
energy density and the energy balance in braneworld models on the AdS bulk
were discussed in \cite{Saha04b}.

For the regularization of the divergent expression in the right-hand side of
(\ref{Evac}) we use the generalized zeta function method \cite{Eliz94}. This
method has been widely used in the evaluation of the Casimir energy for
various bulk and boundary geometries \cite{Most97}, in particular, in
braneworld models \cite{Fabi00}-\cite{Garr03}. Let us introduce the zeta
function related to (\ref{Evac}):%
\begin{equation}
\zeta (s)=\frac{\left( D-1\right) (4\pi )^{(1-D)/2}}{\Gamma ((D-1)/2)}\mu
^{s+1}\sum_{n=1}^{\infty }\int_{0}^{\infty }dk\,\frac{k^{D-2}}{\left(
\lambda _{\nu ,n}^{2}/z_{1}^{2}+k^{2}\right) ^{s/2}},  \label{zetas}
\end{equation}%
considered as a function of the complex variable $s$. The parameter $\mu $
has dimension of mass and is introduced to keep the dimension of the
right-hand side. For the evaluation of the vacuum energy we need the
analytical continuation of $\zeta (s)$ at the point $s=-1$. After the
integration over $k$, (\ref{zetas}) is presented in terms of the partial
zeta function $\zeta _{p}\left( s\right) =\sum_{n=1}^{\infty }\lambda _{\nu
,n}^{-s}$. The corresponding formula reads
\begin{equation}
\zeta (s)=\frac{(D-1)\Gamma ((s+1-D)/2)}{2(4\pi )^{(D-1)/2}\Gamma
(s/2)z_{1}^{D}}(\mu z_{1})^{s+1}\zeta _{p}\left( s+1-D\right) \mathbf{.}
\label{zetas2}
\end{equation}%
Hence, the problem is reduced to the analytic continuation of the partial
zeta function $\zeta _{p}\left( s\right) $ at the point $s=-D$.

In the cases $\nu =\pm 1/2$ the function $g_{\nu ,\nu }(\lambda
z_{1},\lambda z)$ is given by (\ref{g12}) and the eigenvalues $\lambda _{\nu
,n}$ are simplified to (\ref{lamn}). The corresponding partial zeta function
$\zeta _{p}\left( s\right) $ is expressed in terms of the Riemann zeta
function $\zeta _{R}\left( x\right) $. For the function (\ref{zetas2}) one
finds%
\begin{equation}
\zeta (s)=\frac{(D-1)\Gamma ((D-s)/2)\mu ^{s+1}}{2^{D}\pi ^{D/2}\Gamma
(s/2)\left( z_{2}-z_{1}\right) ^{D-1-s}}\zeta _{R}(D-s)\mathbf{.}
\label{zetasSp}
\end{equation}%
In deriving this expression we have used the relation%
\begin{equation}
\pi ^{x+1/2}\Gamma \left( -x/2\right) \zeta _{R}(-x)=\Gamma ((x+1)/2)\zeta
_{R}(x+1),  \label{relZet}
\end{equation}%
for the product of the gamma and Riemann zeta functions. The expression (\ref%
{zetasSp}) is finite at the physical point $s=-1$ and for the vacuum energy
we get%
\begin{equation}
\mathcal{E}_{\mathrm{(II)}}=-(D-1)\frac{\Gamma ((D+1)/2)\zeta _{R}(D+1)}{%
\left( 4\pi \right) ^{(D+1)/2}\left( z_{2}-z_{1}\right) ^{D}}.  \label{Ensp}
\end{equation}%
The special cases under consideration are realized for $D=3$:%
\begin{equation}
\mathcal{E}_{\mathrm{(II)}}=-\frac{\pi ^{2}}{720\left( z_{2}-z_{1}\right)
^{3}},  \label{Ensp2}
\end{equation}%
for both the boundary conditions (\ref{BC}) and (\ref{BC2}). This coincides
with the $D=3$ Casimir energy for plates in the Minkowski bulk with
separation $z_{2}-z_{1}$. In the case of boundary condition (\ref{BC2}) one
has $\nu =1/2$ for $D=5$ and the corresponding energy is given by
\begin{equation}
\mathcal{E}_{\mathrm{(II)}}=-\frac{\pi ^{3}}{7560\left( z_{2}-z_{1}\right)
^{5}}.  \label{Ensp3}
\end{equation}

Now we return to the general case of $\nu $. The partial zeta function is
presented as the contour integral
\begin{equation}
\zeta _{p}\left( s\right) =\frac{1}{2\pi i}\int_{C}\frac{du}{u^{s}}%
\,\partial _{u}\ln \left[ g_{\nu ,\nu }(u,u\eta )\right] ,  \label{zetanint1}
\end{equation}%
in the complex plane $u$. The closed counterclockwise contour $C$ consists
of large semicircle $C_{R}$ in the right half-plane, with the center at $u=0$
and with the radius $R$ tending to infinity, and a straight part that
coincides with the imaginary axis. The point $u=0$ is avoided by semicircle $%
C_{r}$ in the right half-plane with small radius $r$. For $\mathrm{Re}\,s<2$
the contribution from the latter contour to the integral in (\ref{zetanint1}%
) vanishes in the limit $r\rightarrow 0$. The integral is decomposed as
\begin{eqnarray}
\zeta _{p}\left( s\right) &=&\frac{\eta ^{s}}{2\pi i}\int_{C}\frac{du}{u^{s}}%
\,\partial _{u}\ln \left[ u^{-\nu }J_{\nu }(u)\right] +\frac{1}{2\pi i}%
\sum_{j=1,2}\int_{C_{j}}\frac{du}{u^{s}}\,\,\partial _{u}\ln \left[ u^{\nu
}H_{\nu }^{(j)}(u)\right]  \notag \\
&&+\frac{1}{2\pi i}\sum_{j=1,2}\int_{C_{j}}\frac{du}{u^{s}}\,\,\partial
_{u}\ln \left[ 1-\frac{J_{\nu }(u)H_{\nu }^{(j)}(\eta u)}{H_{\nu
}^{(j)}(u)J_{\nu }(\eta u)}\right] ,  \label{zetap2}
\end{eqnarray}%
where $C_{1}$ and $C_{2}$ are the parts of the contour $C$ in upper and
lower half-planes, $H_{\nu }^{(1,2)}(x)$ are the Hankel functions and $\eta $
is defined by (\ref{eta}). In the integrals over the imaginary axis we
introduce the modified Bessel functions. Under the condition $\mathrm{Re}%
\,s>1$, the integrals over the circular parts of the contour $C$ tend to
zero in the limit $R\rightarrow \infty $ and we obtain the integral
representation of the function $\zeta _{p}\left( s\right) $. Substituting in
(\ref{zetas2}), for the zeta function we get%
\begin{eqnarray}
\zeta (s) &=&\frac{(D-1)(4\pi )^{(1-D)/2}\mu ^{s+1}}{2\Gamma (s/2)\Gamma
((D+1-s)/2)}\int_{0}^{\infty }dx\,x^{D-1-s}\partial _{x}  \notag \\
&&\times \left\{ \ln \left[ x^{-\nu }I_{\nu }(xz_{2})\right] +\ln \left[
x^{\nu }K_{\nu }(xz_{1})\right] +\ln \left[ 1-\frac{I_{\nu }(xz_{1})K_{\nu
}(xz_{2})}{K_{\nu }(xz_{1})I_{\nu }(xz_{2})}\right] \right\} .
\label{zetas3}
\end{eqnarray}%
This integral representation is valid in the range $D<$ $\mathrm{Re}\,s<D+1$%
. We need the analytical continuation of this expression to the point $s=-1$.

The part of the zeta function with the last term in figure braces of (\ref%
{zetas3}) is finite at $s=-1$. We will denote the corresponding contribution
to the vacuum energy by $\Delta \mathcal{E}_{\mathrm{(II)}}$. After
integration by parts it is presented in the form%
\begin{equation}
\Delta \mathcal{E}_{\mathrm{(II)}}=\frac{D-1}{(4\pi )^{D/2}\Gamma (D/2)}%
\int_{0}^{\infty }dx\,x^{D-1}\ln \left[ 1-\frac{I_{\nu }(xz_{1})K_{\nu
}(xz_{2})}{K_{\nu }(xz_{1})I_{\nu }(xz_{2})}\right] .  \label{DelE}
\end{equation}%
Now, comparing with (\ref{fja}), we see that the Casimir interaction forces
per unit surface acting on the plates are related to the energy (\ref{DelE})
by the formula%
\begin{equation}
f_{j}^{\mathrm{(int)}}=-(z_{j}{}/\alpha )^{D+1}\frac{\partial }{\partial
z_{j}}\Delta \mathcal{E}_{\mathrm{(II)}}.  \label{fjDelE}
\end{equation}%
This relation provides an alternative way for the evaluation of the Casimir
forces acting on the plates. Note that the main part of previous
investigations of the Casimir effect on the AdS bulk follow this procedure.

\subsection{Zeta function in the geometry with a single plate}

In the limit $z_{1}\rightarrow 0$ the part of the zeta function $\zeta (s)$
with the last two terms in figure braces of (\ref{zetas3}) tends to zero.
Based on this, the part with the first term can be interpreted as the zeta
function for the vacuum energy in the region $0\leq z\leq z_{2}$ for the
geometry of a single plate at $z=z_{2}$. This can also be seen by direct
evaluation. Indeed, for a single plate at $z=z_{j}$ the mode functions for
the vector potential in the region $0\leq z\leq z_{j}$ have the form (\ref%
{Modes}) with $C_{2}=0$. From the boundary conditions it follows that the
eigenvalues of the quantum number $\lambda $ are solutions of the equation $%
J_{\nu }(\lambda z_{j})=0$ with the same $\nu $ as in (\ref{lameig}). The
total energy of the vacuum per unit coordinate volume along the parallel
directions is given by (\ref{Evac}), with $z_{1}$ replaced by $z_{j}$, where
now $\lambda _{\nu ,n}$ is the $n$-th positive zero of the function $J_{\nu
}(x)$. Introducing the related zeta function $\zeta _{\mathrm{(I)}}(s,z_{j})$
similar to (\ref{zetas}), by transformations like those presented above, the
following integral representation is obtained%
\begin{equation}
\zeta _{\mathrm{(I)}}(s,z_{j})=\frac{(D-1)(4\pi )^{(1-D)/2}\left( \mu
z_{j}\right) ^{s+1}}{2\Gamma (s/2)\Gamma ((D+1-s)/2)z_{j}^{D}}%
\int_{0}^{\infty }dx\,x^{D-1-s}\partial _{x}\ln \left[ x^{-\nu }I_{\nu }(x)%
\right] ,  \label{ZetL}
\end{equation}%
with the range of validity $D<\mathrm{Re\,}s<D+1$.

The analytical continuation of (\ref{ZetL}) to the physical point $s=-1$ is
done by the method widely discussed in the context of the Casimir effect. As
the first step we decompose the integral into integrals over the regions $%
x\in \lbrack 0,1]$ and $x\in \lbrack 1,\infty )$. The part of the zeta
function with the first integral is finite at $s=-1$. For the analytic
continuation of the part with the second integral we subtract and add in the
integrand the first $N$ terms of the corresponding asymptotic expansion for
large values of $x$. This gives%
\begin{eqnarray}
\zeta _{\mathrm{(I)}}(s,z_{j}) &=&\frac{(D-1)(4\pi )^{(1-D)/2}\left( \mu
z_{j}\right) ^{s+1}}{2\Gamma (s/2)\Gamma ((D+1-s)/2)z_{j}^{D}}\left\{
\int_{0}^{1}dx\,x^{D-1-s}\partial _{x}\ln \left[ x^{-\nu }I_{\nu }(x)\right]
\right.  \notag \\
&&\left. +\int_{1}^{\infty }dx\,x^{D-1-s}\left[ \partial _{x}\ln \left(
x^{-\nu }I_{\nu }(x)\right) -\sum_{k=0}^{N}\frac{b_{I,k}}{x^{k}}\right]
-\sum_{k=0}^{N}\,\frac{b_{I,k}}{D-s-k}\right\} ,  \label{ZetL1}
\end{eqnarray}%
where $b_{I,k}$ are the coefficients of the asymptotic expansion of the
function $\partial _{x}\ln \left( x^{-\nu }I_{\nu }(x)\right) $ for large $x$%
. For those coefficients one has%
\begin{equation}
b_{I,0}=1,\;b_{I,1}=-\nu -1/2,\;b_{I,k}=-\left( k-1\right) \beta
_{k-1},\;k\geq 2.  \label{bk}
\end{equation}%
Here, $\beta _{k}$ are defined by the relation
\begin{equation}
\ln \left( \sum_{k=0}^{\infty }\frac{\alpha _{k}}{x^{k}}\right)
=\sum_{k=1}^{\infty }\frac{\beta _{k}}{x^{k}},  \label{betk}
\end{equation}%
and%
\begin{equation}
\alpha _{k}=\frac{(-1)^{k}\Gamma \left( \nu +k+1/2\right) }{2^{k}k!\Gamma
\left( \nu -k+1/2\right) }.  \label{alfk}
\end{equation}%
are the coefficients of the asymptotic expansion of the function $\sqrt{2\pi
x}e^{-x}I_{\nu }(x)$ (the expressions for the first six coefficients $\beta
_{k}$ are given, for example, in \cite{Flac01}).

For $N>D$, in (\ref{ZetL1}) the integral over $[1,\infty )$ is finite at $%
s=-1$. The only singularity at this physical point comes from the simple
pole corresponding to the term $k=D+1$ of the last sum in figure braces of (%
\ref{ZetL1}). Denoting by $\zeta _{\mathrm{(I)}}^{(p)}(s,z_{j})$ and $\zeta
_{\mathrm{(I)}}^{(f)}(z_{j})$ the pole and finite parts of the zeta
function, from (\ref{ZetL1}) one gets%
\begin{equation}
\zeta _{\mathrm{(I)}}^{(p)}(s,z_{j})=-\frac{(D-1)b_{I,D+1}}{D(4\pi
)^{D/2}\Gamma (D/2)z_{j}^{D}}\frac{1}{s+1}.  \label{ZetLp}
\end{equation}%
The finite part of the zeta function gives the finite part of the vacuum
energy induced by the brane $z=z_{j}$ in the region $0\leq z\leq z_{j}$:%
\begin{eqnarray}
\mathcal{E}_{\mathrm{(I)}}^{(f)}(z_{j}) &=&\zeta _{\mathrm{(I)}%
}^{(f)}(z_{j})=-\frac{(D-1)\pi ^{-D/2}}{2^{D}D\Gamma (D/2)z_{j}^{D}}\left\{
\mathbf{\,}\int_{0}^{1}dx\,x^{D}\partial _{x}\ln \left( \frac{I_{\nu }(x)}{%
x^{\nu }}\right) \right.  \notag \\
&&+\int_{1}^{\infty }dx\,x^{D}\left[ \partial _{x}\ln \left( \frac{I_{\nu
}(x)}{x^{\nu }}\right) -\sum_{k=0}^{N}\frac{b_{I,k}}{x^{k}}\right]  \notag \\
&&\left. +b_{I,D+1}\left[ \ln \left( \mu z_{j}\right) -\psi \left( \frac{D}{2%
}+1\right) -\psi \left( -\frac{1}{2}\right) \right] +\sideset{}{'}{\sum}%
_{k=0}^{N}\,\frac{b_{I,k}}{k-D-1}\right\} ,  \label{ZetLf}
\end{eqnarray}%
where $\psi (x)=\Gamma ^{\prime }(x)/\Gamma (x)$ is the digamma function and
the prime on the summation sign means that the term $k=D+1$ is excluded from
the sum.

In order to provide a physical interpretation for the second term in figure
braces of (\ref{zetas3}) let us consider the limit $z_{2}\rightarrow \infty $%
. The contributions with the first and third terms tend to zero. This allows
us to interpret the function
\begin{equation}
\zeta _{\mathrm{(III)}}(s,z_{j})=\frac{(D-1)(4\pi )^{(1-D)/2}\left( \mu
z_{j}\right) ^{s+1}}{2\Gamma (s/2)\Gamma ((D+1-s)/2)z_{j}^{D}}%
\int_{0}^{\infty }dx\,x^{D-1-s}\partial _{x}\ln \left[ x^{\nu }K_{\nu }(x)%
\right] ,  \label{ZetR}
\end{equation}%
as the zeta function for the region $z_{j}\leq z<\infty $ in the geometry of
a single plate at $z=z_{j}$. The validity range of this representation is
the same as that for (\ref{ZetL}). The analytical continuation of the zeta
function (\ref{ZetR}) is similar to that for the function (\ref{ZetL}). The
corresponding pole part is presented as%
\begin{equation}
\zeta _{\mathrm{(III)}}^{(p)}(s,z_{j})=-\frac{(D-1)b_{K,D+1}}{D(4\pi
)^{D/2}\Gamma (D/2)z_{j}^{D}}\frac{1}{s+1}.  \label{ZetRp}
\end{equation}%
The expression for the finite part of the vacuum energy, induced by the
plate at $z=z_{j}$ in the region $z_{j}\leq z<\infty $, reads
\begin{eqnarray}
\mathcal{E}_{\mathrm{(III)}}^{(f)}(z_{j}) &=&\zeta _{\mathrm{(III)}%
}^{(f)}(z_{j})=-\frac{(D-1)\pi ^{-D/2}}{2^{D}D\Gamma (D/2)z_{j}^{D}}\left\{
\mathbf{\,}\int_{0}^{1}dx\,x^{D}\partial _{x}\ln \left( x^{\nu }K_{\nu
}(x)\right) \right.  \notag \\
&&+\int_{1}^{\infty }dx\,x^{D}\left[ \partial _{x}\ln \left( x^{\nu }K_{\nu
}(x)\right) -\sum_{k=0}^{N}\frac{b_{K,k}}{x^{k}}\right]  \notag \\
&&\left. +b_{K,D+1}\left[ \ln \left( \mu z_{j}\right) -\psi \left( \frac{D}{2%
}+1\right) -\psi \left( -\frac{1}{2}\right) \right] +\sideset{}{'}{\sum}%
_{k=0}^{N}\,\frac{b_{K,k}}{k-D-1}\right\} ,  \label{ZetRf}
\end{eqnarray}%
where%
\begin{equation}
b_{K,0}=-1,\;b_{K,1}=\nu -1/2,\;b_{K,k}=\left( -1\right) ^{k}\left(
k-1\right) \beta _{k-1},\;k\geq 2.  \label{bkK}
\end{equation}%
In deriving (\ref{ZetRp}) and (\ref{ZetRf}) we have used the asymptotic
expansion $\sqrt{2x/\pi }e^{x}K_{\nu }(x)=\sum_{k=0}^{\infty }(-1)^{k}\alpha
_{k}/x^{k}$ for the Macdonald function. Note that $b_{K,k}=-(-1)^{k}b_{I,k}$
for $k\geq 2$.

\subsection{Total zeta function and the vacuum energy}

Based on (\ref{ZetLp}), (\ref{ZetLf}), (\ref{ZetRp}), and (\ref{ZetRf}), for
the pole part of the zeta function in the region between the plates we find%
\begin{equation}
\zeta ^{(p)}(s)=\zeta _{\mathrm{(III)}}^{(p)}(s,z_{1})+\zeta _{\mathrm{(I)}%
}^{(p)}(s,z_{2}).  \label{Zetp}
\end{equation}%
The pole terms in the right-hand side come from the single plate
contributions. The finite part of the vacuum energy in the region between
the plates is given by
\begin{equation}
\mathcal{E}_{\mathrm{(II)}}^{(f)}=\mathcal{E}_{\mathrm{(III)}}^{(f)}(z_{1})+%
\mathcal{E}_{\mathrm{(I)}}^{(f)}(z_{2})+\Delta \mathcal{E}_{\mathrm{(II)}}.
\label{Ef}
\end{equation}%
The divergent part of the vacuum energy coming from the pole term is
absorbed by renormalizing the cosmological constants on the plates. Indeed,
if $g_{(j)ik}$ is the induced metric tensor on the plate at $z=z_{j}$ and $%
R_{(j)}$ is the corresponding Ricci scalar, then the surface action for that
plate is given by
\begin{equation}
S_{(j)}=\frac{\left( \alpha /z_{j}\right) ^{D}}{16\pi G_{(j)}}\int d^{D}xdz\,%
\sqrt{|g_{(j)}|}\left[ R_{(j)}-2\Lambda _{(j)}\right] \delta (z-z_{j}),
\label{Sj}
\end{equation}%
where $\Lambda _{(j)}$ is the cosmological constant on the plate. As seen,
the pole term in the vacuum energy has the same dependence on $z_{j}$ ($%
z_{j}^{-D}$) as the part in the action with the cosmological constant and
can be absorbed by the renormalization of the latter. This point has been
widely discussed in \cite{Fabi00}-\cite{Garr03} in the context of
Randall-Sundrum braneworlds.

Having the zeta functions in separate regions $0\leq z\leq z_{1}$, $%
z_{1}\leq z\leq z_{2}$ and $z_{2}\leq z<\infty $ we can find the total zeta
function:
\begin{eqnarray}
\zeta _{\mathrm{tot}}(s) &=&\frac{(D-1)(4\pi )^{(1-D)/2}\mu ^{s+1}}{2\Gamma
(s/2)\Gamma ((D+1-s)/2)}\int_{0}^{\infty }dx\,x^{D-1-s}\partial _{x}  \notag
\\
&&\times \left\{ \sum_{j=1,2}\ln \left[ I_{\nu }(xz_{j})K_{\nu }(xz_{j})%
\right] +\ln \left[ 1-\frac{I_{\nu }(xz_{1})K_{\nu }(xz_{2})}{K_{\nu
}(xz_{1})I_{\nu }(xz_{2})}\right] \right\} .  \label{Zetatot}
\end{eqnarray}%
The parts with the separate terms of the sum over $j$ are the total zeta
functions in the problems with single plates at $z=z_{j}$. For the
corresponding pole part one has
\begin{equation}
\zeta _{\mathrm{tot}}^{(p)}(s)=-\frac{1+(-1)^{D}}{s+1}\sum_{j=1,2}\frac{%
(D-1)b_{I,D+1}}{D(4\pi )^{D/2}\Gamma (D/2)z_{j}^{D}},  \label{Aetatotp}
\end{equation}%
where we have used the relation $b_{K,D+1}=(-1)^{D}b_{I,D+1}$. It is the sum
of the pole parts for single plate geometries. For odd values of the spatial
dimension $D$ the total pole part is zero and the finite part does not
depend on the renormalization scale $\mu $. In this case the total vacuum
energy is unambiguously defined and is given by
\begin{equation}
\mathcal{E}=\sum_{j=1,2}\mathcal{E}_{j}+\Delta \mathcal{E}_{\mathrm{(II)}},
\label{Etot}
\end{equation}%
where%
\begin{eqnarray}
\mathcal{E}_{j} &=&-\frac{(D-1)\pi ^{-D/2}}{2^{D}D\Gamma (D/2)z_{j}^{D}}%
\left\{ \mathbf{\,}\frac{1}{D}-\sum_{l=1}^{N_{1}}\,\frac{4l\beta _{2l}}{2l-D}%
+\int_{0}^{1}dx\,x^{D}\partial _{x}\ln \left( I_{\nu }(x)K_{\nu }(x)\right)
\right.  \notag \\
&&\left. +\int_{1}^{\infty }dx\,x^{D-1}\left[ x\partial _{x}\ln \left(
I_{\nu }(x)K_{\nu }(x)\right) +1+4\sum_{l=1}^{N_{1}}\frac{l\beta _{2l}}{%
x^{2l}}\right] \right\} ,  \label{Ej}
\end{eqnarray}%
with $N_{1}\geq (D+1)/2$, is the total vacuum energy for the geometry of a
single boundary at $z=z_{j}$. For $D=3$, $\nu =D/2-1=1/2$ one finds%
\begin{equation}
\mathcal{E}_{j}=-\frac{\pi ^{2}}{720z_{j}^{3}},  \label{EjD3}
\end{equation}%
and for $D=3$, $\nu =D/2-2=-1/2$ we get%
\begin{equation}
\mathcal{E}_{j}=\frac{7\pi ^{2}}{5760z_{j}^{3}}.  \label{EjD3b}
\end{equation}%
In table \ref{tab1} we present the Casimir energy $\mathcal{E}_{j}$ in a
single plate geometry for odd spatial dimensions and for boundary conditions
(\ref{BC}) and (\ref{BC2}). As seen, depending on the boundary condition and
on the spatial dimension the energy can be either positive or negative.

\begin{table}[tbp]
\caption{The Casimir energies for a single plate in different numbers of odd
spatial dimensions for the boundary conditions (\protect\ref{BC}) and (%
\protect\ref{BC2}).}
\label{tab1}\centering
\begin{tabular}{|c|c|c|c|c|}
\hline\hline
$D$ & $3$ & $5$ & $7$ & $9$ \\ \hline
$z_{j}^{D}\mathcal{E}_{j}$, BC (\ref{BC}) & $-0.01371$ & $0.01269$ & $%
-0.01640$ & $0.02701$ \\ \hline
$z_{j}^{D}\mathcal{E}_{j}$, BC (\ref{BC2}) & $0.01199$ & $-0.00410$ & $%
0.00312$ & $-0.00378$ \\ \hline
\end{tabular}%
\end{table}

The $z$-component of the force acting per unit surface of the plate at $%
z=z_{j}$ can be evaluated by using the formula%
\begin{equation}
f_{j}=-(z_{j}{}/\alpha )^{D+1}\frac{\partial }{\partial z_{j}}\mathcal{E}.
\label{fj2}
\end{equation}%
Based on the decomposition (\ref{Etot}) of the vacuum energy the force is
presented in the form (\ref{fj}), where the interaction part $f_{j}^{\mathrm{%
(int)}}=(-1)^{j}p_{j}^{\mathrm{(int)}}$ is obtained from the contribution $%
\Delta \mathcal{E}_{\mathrm{(II)}}$ in (\ref{Etot}) (see (\ref{fjDelE})).
The self-action part of the force comes from the single plate energies:%
\begin{equation}
f_{j}^{\mathrm{(s)}}=-(z_{j}{}/\alpha )^{D+1}\frac{\partial }{\partial z_{j}}%
\mathcal{E}_{j}=\frac{Dz_{j}^{D}}{\alpha ^{D+1}}\mathcal{E}_{j}.  \label{fjs}
\end{equation}%
Comparing with (\ref{Ej}) we see that the self-action force does not depend
on the location of the plate and it is the same for both plates. The
corresponding values of $z_{j}^{D}\mathcal{E}_{j}$ for odd number of spatial
dimensions are given in table \ref{tab1}. The force acting on the plate is
attractive with respect to the AdS boundary for $\mathcal{E}_{j}<0$ and
repulsive for $\mathcal{E}_{j}>0$. It has opposite signs for the boundary
conditions (\ref{BC}) and (\ref{BC2}).

\section{Applications to $Z_{2}$-symmetric braneworld models}

\label{sec:Branes}

Higher-dimensional braneworld models of the Randall-Sundrum type \cite%
{Rand99} are among modern developments in theoretical physics where the AdS
spacetime plays a central role. The original model with two branes is
formulated on a slice of the $D=4$ AdS bulk with the extra dimension $x^{4}$
compactified on a $S^{1}/Z_{2}$ orbifold. The branes are located at the
fixed points of the orbifold. Considering general number of spatial
dimensions $D$, the metric tensor in terms of the radial coordinate $y$ is
given by $g_{\mu \nu }=\mathrm{diag}(e^{-2|y|/\alpha }\eta _{ik},-1)$ with
the fixed points $y=0$ and $y=a$ and with $-a\leq y\leq a$. The two regions $%
y<0$ and $y>0$ are identified by the $Z_{2}$ symmetry. The boundary
conditions for fields propagating in the bulk are dictated by that symmetry
(see, for example, the discussion in \cite{Gher00,Chan05} for the case $D=4$%
).

In fact, one has two reflection symmetries. The first one corresponds to the
reflection with respect to the brane at $y=0$ with $y\rightarrow -y$ and the
second one is the symmetry under the reflection with respect to the brane at
$y=a$ with $y-a\rightarrow a-y$. For fields even under both these
reflections the mode functions are given by (\ref{Modes}) with $z=\alpha
e^{|y|/\alpha }$. The corresponding boundary conditions are obtained by the
integration of the field equation near the points $y=0$ and $y=a$. The
conditions on the radial function are reduced to the constraints $\partial
_{D}A_{(\beta )l}=0$ for $y=0,a$, and correspond to the boundary condition (%
\ref{BC2}) discussed in previous sections. For odd fields the modes are
obtained from (\ref{Modes}) taking $z=\alpha e^{|y|/\alpha }$ and adding an
additional coefficient $\mathrm{sgn}(y)$. In this case from the continuity
of the modes at the locations of the branes we get the boundary conditions $%
A_{(\beta )l}=0$ for $y=0,a$, that correspond to the condition (\ref{BC}).
From this consideration it follows that the modes for massless vector fields
in higher-dimensional generalization of the Randall-Sundrum models are given
by (\ref{Modes2}) with $\nu =D/2-1$ for odd fields under the $Z_{2}$
symmetry and $\nu =D/2-2$ for even fields. The normalization constants in
both these cases are given by (\ref{C}) with an additional factor 1/2 in the
right-hand side. The latter difference is related to the fact that now the
integration over $y$ in the normalization condition (\ref{NC}) goes over the
region $[-a,a]$ instead of $[0,a]$ in the problem we have discussed before.
Hence, the expressions for the local characteristics of the electromagnetic
vacuum in $Z_{2}$-symmetric braneworlds, such as the correlators, the VEVs
of the fields squared, of the energy-momentum tensor and the Casimir forces,
are obtained from the expressions given above with an additional factor 1/2
and with $z_{1}=\alpha $, $z_{2}=\alpha e^{a/\alpha }$. For odd and even
fields with respect to the $Z_{2}$ reflection the results for the boundary
conditions (\ref{BC}) and (\ref{BC2}) have to be taken. The expression for
the total Casimir energy in the region between the branes remains the same
(the integration over the region $y\in \lbrack -a,a]$ brings a factor of 2).
The regions I and III we have discussed above are excluded in the braneworld
setup.

We can also consider the case when the $Z_{2}$-parities of the field with
respect to the branes $y=0$ and $y=a$ have opposite signs (for different
combinations of parities in the case of vector fields see \cite{Chan05}).
For example, let us discuss the field odd under the reflection with respect
to the brane $y=0$ and even under the reflection with respect to the second
brane at $y=a$. Now, the field obeys the boundary condition $A_{(\beta )l}=0$
for $y=0$ and the condition $\partial _{D}A_{(\beta )l}=0$ for $y=a$. In the
region $0\leq y\leq a$ the mode functions are given by (\ref{Modes}). From
the boundary condition at $y=0$ it follows that $c_{2}/c_{1}=-J_{D/2-1}(%
\lambda z_{1})/Y_{D/2-1}(\lambda z_{1})$. For the corresponding modes we get
\begin{equation}
A_{(\beta )\mu }=C\epsilon _{(\sigma )\mu }z^{D/2-1}g_{D/2-1,D/2-1}(\lambda
z_{1},\lambda z)e^{ik_{l}x^{l}}.  \label{Modes3}
\end{equation}%
The boundary condition on the brane $y=a$ leads to the equation $%
g_{D/2-1,D/2-2}(\lambda z_{1},\lambda z_{2})=0$ for the eigenvalues of the
quantum number $\lambda $. The normalization coefficient is obtained from (%
\ref{NC}) with the integration over the region $y\in \lbrack -a,a]$ and is
given by%
\begin{equation}
|C|^{2}=\frac{\pi ^{2}\alpha ^{3-D}\lambda _{n}^{2}}{4\left( 2\pi \right)
^{D-2}\omega z_{1}^{2}}\left[ \frac{J_{D/2-1}^{2}(\lambda _{n})}{%
J_{D/2-2}^{2}(\lambda _{n}\eta )}-1\right] ^{-1},  \label{Cbr3}
\end{equation}%
with $\lambda _{n}$ being the eigenvalues for $\lambda z_{1}$. The
investigation of the VEVs is similar to that we have described in the
previous sections. The summation formula for the series in the mode sums
over the eigenvalues of $\lambda $ is given in \cite{Saha87,Saha08Book}.
Note that we could consider the analog of the problems discussed in the
previous sections with the boundary condition (\ref{BC}) on the plate $%
z=z_{1}$ and the condition (\ref{BC2}) on the plate $z=z_{2}$. The
corresponding mode functions in the region $z_{1}\leq z\leq z_{2}$ are given
by (\ref{Modes3}) with an additional factor 2 in the expression (\ref{Cbr3})
for the normalization coefficient.

\section{Summary}

\label{sec:Conc}

We have discussed the effects of two parallel plates in AdS spacetime on the
properties of the electromagnetic vacuum in an arbitrary number of spatial
dimensions. The plates are parallel to the AdS boundary and two types of
boundary conditions were discussed on them. The first one is the
generalization of the perfect conductor boundary condition for an arbitrary
number of spatial dimensions and the second one corresponds to the confining
boundary condition used in quantum chromodynamics to confine gluons. In the
model under consideration the properties of the electromagnetic vacuum are
encoded in two-point functions and we have started the consideration from
the two-point function of the vector potential. It is presented in the form
of the sum over complete set of electromagnetic modes. In the region between
the plates, the eigenvalues of the quantum number corresponding to the
direction normal to the plates are roots of the equation (\ref{lameig}). The
application of the summation formula (\ref{SumAbel}) to the corresponding
series allowed us to extract the contribution of the single plate and to
present the second plate induced part in the form that is well adapted for
the evaluation of the VEVs for local observables. Two equivalent
representations of the two-point function for the vector potential are given
by (\ref{AA2}) and (\ref{AA3}). The VEVs of local physical observables are
obtained from the two-point functions for the field tensor and the
expressions for those functions are given in section \ref{sec:2pF}. We have
provided the expressions for both the single plate and the second plate
induced contributions.

As important local characteristics of the vacuum state we have considered
the VEVs of the electric and magnetic fields squared and the photon
condensate. The single plate and the second plate induced contributions to
the VEV of the electric field square are positive for the boundary condition
(\ref{BC}) and negative for the condition (\ref{BC2}). The signs of the
corresponding contributions in the VEV of the magnetic field squared and in
the photon condensate are opposite to that for the electric field. Near the
plates the VEVs are dominated by single plate contributions. In those
regions the effects of the gravity on the VEVs are weak and the leading
terms in the asymptotic expansions over the distance from the plate coincide
with those in the corresponding problem on the Minkwoski bulk. The effects
of the gravity are essential for separations between the plates larger than
the curvature radius of the background geometry. Having the correlators for
the field tensor one can evaluate the Casimir-Polder forces acting on a
polarizable particle. As an illustration we have considered the simplest
case of isotropic polarizability in the static limit. Near the plate, the
Casimir-Polder force is attractive/repulsive with respect to that plate for
the boundary condition (\ref{BC})/(\ref{BC2}).

Similar investigations for the VEV of the energy-momentum tensor are
presented in section \ref{sec:emt}. The off-diagonal components vanish and
the diagonal components are decomposed as (\ref{Tmu}), (\ref{TD}), with the
single plate contributions given by (\ref{Tj1}). The vacuum stresses along
the directions parallel to the plates are equal to the energy density. We
have checked that the boundary induced VEV in the energy-momentum tensor
obeys the covariant continuity equation and its trace is expressed in terms
of the photon condensate as (\ref{Trace}). For $D\geq 4$ the boundary induced
contribution in the vacuum energy density is negative for the condition (\ref%
{BC}) and positive for (\ref{BC2}). In $D=3$ spatial dimensions the
electromagnetic field is conformally invariant and the vacuum
energy-momentum tensor in the region II is given by simple expressions (\ref%
{TmuD3}). The latter is the same for the boundary conditions (\ref{BC}) and (%
\ref{BC2}). For $D=3$ the boundary induced VEV in the energy-momentum tensor
vanishes in the region III. In the region I the vacuum energy-momentum
tensor is given by (\ref{TmuD3L}) for the condition (\ref{BC}) and by (\ref%
{TDDcL}) for (\ref{BC2}). Those expressions are different and that is a
consequence of different interactions with the AdS boundary.

Based on the expressions for the normal stress we have investigated the
Casimir forces acting on the plates. They are decomposed into the
self-action and interaction contributions. The first parts come from the
single plate contributions in the normal stress and for $D\neq 3$ require an
additional renormalization because of the surface divergences. The
interaction parts in the vacuum pressures on the plates are given by (\ref%
{p12}) and they are attractive for both the boundary conditions (\ref{BC})
and (\ref{BC2}). An alternative expressions are given by (\ref{fja}). For $%
D=3$ the self-action forces are finite and the total forces per unit surface
of the plates are presented as (\ref{f12BC}) and (\ref{f2BC2}). The force on
the right plate is directed toward the AdS boundary for both the boundary
conditions. The force on the left plate for the condition (\ref{BC2}) is
directed toward the AdS horizon. For the condition (\ref{BC}) the force on
the left plate is directed towards the AdS horizon for small separations and
towards the AdS boundary for large separations. For general $D$ and at small
separations between the plates, compared to the AdS curvature radius, the
leading terms in the Casimir forces coincide with that for the plates in the
Minkowski bulk. This is a consequence of the fact that at small separations
the contribution of the vacuum fluctuations with small wavelengths dominates
in the vacuum forces and the influence of gravity on those fluctuations is
weak. At separations larger than the AdS curvature scale the gravity
essentially changes the behavior of the Casimir forces: considered as
functions of the interplate separation, the forces decay exponentially in
contrast to the power-law decay for the Minkowski bulk. Another feature
differing the AdS and Minkowski bulks is that the forces on the left and
right plates differ.

Among the interesting directions in the investigations of the Casimir effect
is the relation between the local and global characteristics of the vacuum
state. In section \ref{sec:VacEn}, the total vacuum energy per unit surface
of the plate is considered. For regularization of the corresponding
divergent expression we have used the generalized zeta function approach. By
using the standard technique, an integral representation of the zeta
function is provided for the region between the plates with separated single
plate and interaction contributions. The latter is finite at the physical
point and the analytic continuation is required for single plate parts only.
In order to do that we have employed the well-known procedure from the
theory of the Casimir effect. The pole and finite parts of the zeta
functions are explicitly provided for both regions in the geometry of a
single plate. The pole parts can be absorbed by the renormalization of the
cosmological constant on the plate. Considering the total zeta function for
a single plate as the sum of the zeta functions in separate regions, the
pole parts are canceled in odd numbers of spatial dimensions and the total
vacuum energy, given by (\ref{Ej}), does not depend on the mass scale in the
renromalization procedure. For a single plate, the force per unit surface of
the plate is obtained by differentiation of the Casimir energy and does not
depend on the location of the plate. Depending on the boundary condition and
on the spatial dimension, that force can be either attractive or repulsive
with respect to the AdS boundary (see the numerical data in table \ref{tab1}%
).

The results obtained can be directly used for the investigation of quantum
vacuum effects in $Z_{2}$-symmetric braneworlds of the Randall-Sundrum type.
The boundary conditions on the branes are dictated by the $Z_{2}$ symmetry
and are reduced to the condition (\ref{BC}) for odd fields and to the
condition (\ref{BC2}) for even fields. The expressions for both the local
and global characteristics of the vacuum state are obtained from those we
have discussed. One can consider also the situation when the $Z_{2}$%
-parities of the fields on the branes are different. In this case we have
the condition (\ref{BC}) on one brane and the condition (\ref{BC2}) on the
other. The corresponding mode functions are given by (\ref{Modes3}) with the
normalization constant (\ref{Cbr3}). The evaluation procedure for the VEVs
is similar to that we have described in this paper.

\section*{Acknowledgments}

A.S.K. and H.G.S. were supported by the grant No. 18T-1C355 of the Committee
of Science of the Ministry of Education, Science, Culture and Sport RA.

\end{document}